\documentclass{egpubl}
 
\JournalSubmission    
\usepackage[T1]{fontenc}
\usepackage{dfadobe}  
\usepackage{capt-of}

\usepackage{cite}  
\usepackage{booktabs}
\BibtexOrBiblatex
\electronicVersion
\PrintedOrElectronic
\ifpdf \usepackage[pdftex]{graphicx} \pdfcompresslevel=9
\else \usepackage[dvips]{graphicx} \fi

\usepackage{egweblnk} 
\usepackage{xcolor}
\usepackage{makecell}

\newcommand{\agp}[1]{{\color{black}#1}}

\newcommand{\revision}[1]{{\color{black}#1}}
\newcommand\blfootnote[1]{%
	\begingroup
	\renewcommand\thefootnote{}\footnote{#1}
	\addtocounter{footnote}{-2}
	\endgroup
}

\title[Advances in Data-Driven Analysis and Synthesis of 3D Indoor Scenes]%
      {Advances in Data-Driven Analysis and Synthesis of 3D Indoor Scenes} 

\author[A. Gadi Patil \& S. Gadi Patil \& M.Li \& M.Fisher \& M.Savva \& H.Zhang]
{\parbox{\textwidth}{\centering Akshay Gadi Patil$^{1}$ \qquad
        Supriya Gadi Patil$^{1}$ \qquad
        Manyi Li$^{2 \dagger}$\qquad
        Matthew Fisher$^{3}$ \qquad
        Manolis Savva$^{1}$ \qquad
        Hao Zhang$^{1}$\newline\\
        $^{1}$Simon Fraser University
        \qquad \qquad
        $^{2}$ Shandong University
        \qquad \qquad
        $^{3}$ Adobe Research}
        }

\begin{document}



\twocolumn[{%
\renewcommand\twocolumn[1][]{#1}%
\maketitle
\begin{center}
    \centering
    \includegraphics[width=\linewidth]{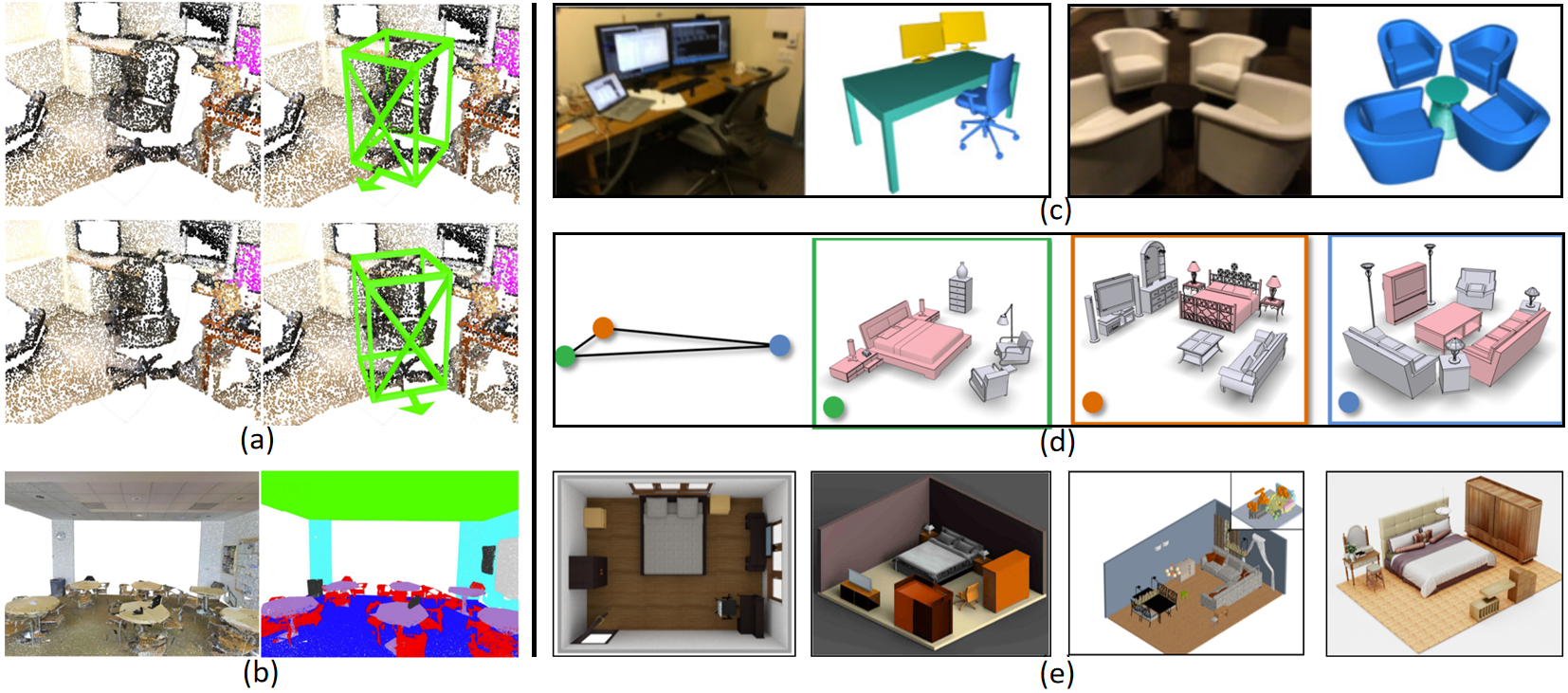}
    \captionof{figure}{A sampler of representative results for different indoor scene modeling tasks surveyed in this report – (a) 3D object detection \cite{yu2022rotationally} \emph{(Section \ref{subsec:obj_det})}, (b) 3D scene segmentation \cite{zhao2021point} \emph{(Section \ref{subsec:3d_ssg})}, (c) scene reconstruction as image-CAD model alignment \cite{gumeli2022roca} \emph{(Section \ref{subsec:scene_recon})}, (d) 3D scene similarity \cite{xu2014organizing} \emph{(Section \ref{subsec:scene_sim})}, and (e) 3D scene synthesis \cite{wang2018deep, li2019grains, yang2021indoor, paschalidou2021atiss} \emph{(Section \ref{sec:synthesis})}. We survey advances in these indoor scene modeling tasks mainly in the realm of 3D geometry. \newline}
    \label{fig:teaser}
\end{center}%
}]

\begin{abstract}
%
\revision{
This report surveys advances in deep learning-based modeling techniques that address four different 3D indoor scene analysis tasks, as well as synthesis of 3D indoor scenes.}
We describe different kinds of representations for indoor scenes, various indoor scene datasets available for research in the aforementioned areas, and discuss notable works employing machine learning models for such scene modeling tasks based on these representations. 
Specifically, we focus on the \emph{analysis} and \emph{synthesis} of 3D indoor scenes. 
With respect to analysis, we focus on four basic scene understanding tasks -- 3D object detection, 3D scene segmentation, 3D scene reconstruction and 3D scene similarity. 
And for synthesis, we mainly discuss neural scene synthesis works, though also highlighting model-driven methods that allow for human-centric, progressive scene synthesis.
We identify the challenges involved in modeling scenes for these tasks and the kind of machinery that needs to be developed to adapt to the data representation, and the task setting in general. For each of these tasks, we provide a comprehensive summary of the state-of-the-art works across different axes such as the choice of data representation, backbone, evaluation metric, input, output etc., providing an organized review of the literature.
Towards the end, we discuss some interesting research directions that have the potential to make a direct impact on the way users interact and engage with these virtual scene models, making them an integral part of the metaverse.

\end{abstract}
\blfootnote{$\dagger$Corresponding Author: manyili@sdu.edu.cn}
\section{Introduction}
\label{sec:intro}

A central goal in computer graphics (CG) is to develop tools for generating real as well as imagined artifacts and environments, such as 3D objects and scenes. 
The pursuit of this goal has been revived in the past decade with a remarkable development in computing technology, including but not limited to, hardware, compute and machine learning algorithms. 
Specifically, the dawn of the big-data era in visual computing coupled with the fast assimilation of deep learning technology has pushed the frontiers of CG research, especially in the realm of content creation and understanding. In this report, we narrow down the focus of the word ``content", to simply refer to 3D indoor scenes. 
%

\revision{In real life, an indoor scene is physically realized by a sequential arrangement of objects in a region-bounded indoor space.} The ubiquity of 3D indoor scenes in real life, has placed an increasing demand for simulations in a wide variety of applications, ranging from  AR/VR, video games, and indoor navigation, to creating virtual runs for AI agents that live and interact in those environments. These indoor scenes are characterized by their constituent elements – 3D object models laid out in a spatially constrained manner. These objects need to be \emph{held together} in some form for functional reasoning and/or contextual interpretation based on the intended human activity.

 To vividly simulate such indoor environments, one needs access to a repository of 3D object models, and possess familiarity with not-so-easy 3D modeling tools. A proxy to this would be to collect large quantities of real-world scenes through acquisition devices (stored as sequences of RGB-D image frames, which can then be converted to a 3D point cloud) and perform object reconstruction that adheres to the captured scene layout. This alternative has its own unavoidable limitations -- captured scenes have inherent errors arising out of sensor limitations that need to be processed before deploying for downstream scene modeling tasks, and 3D reconstructions at both the object level and arrangement level are poor. This premise, then, opens up a multitude of research possibilities in 3D indoor scene modeling, with a \emph{analysis-for-synthesis} theme, keeping in mind the overall goal of content creation.

The first step in reconstructing an acquired 3D scene is to understand its composition, which reduces to localizing constituent 3D objects. Given a large collection of such real-world scans, algorithms can be developed that can learn the occurrence and placement patterns of prominent/all objects, leveraging the localization module. These priors can be used to generate more of such scene layouts, tackling the content creation bottleneck at the arrangement level. Though object-level reconstruction from images/scans is a challenging task, existing approaches could be borrowed to roughly visualize the underlying objects.
In addition, semantic scene segmentation can complement 3D object localization (and vice-versa) in heavily occluded scenes at the object level, leading to better scene reconstruction. The knowledge gained during these analysis tasks can help in generating diverse scenes. We, therefore, focus on modeling scenes in the context of both \emph{analysis} and \emph{synthesis} tasks.

\subsection{Related Surveys}
Our focus is on data-driven modeling of indoor scenes, which includes both \emph{analysis} and \emph{synthesis} of scenes, irrespective of their representation. In the past, \cite{pintore2020state} focused on the structured reconstruction of 3D indoor scenes, and \cite{chaudhuri2020learning} focused on generative models for 3D structures, which partly covers 3D indoor scenes as applications to presented approaches in different papers surveyed. Both reports focus on structural methods, one on reconstruction and the latter on generation, respectively. Our report differs from the two in the sense that it is not confined to structured modalities, and includes different aspects of scene analysis, going beyond reconstruction. As well, for scene generation, we focus mainly on neural generation, though also highlighting model-driven methods that allow for human-centric, progressive scene synthesis. With a mix of historical and contemporary works, we provide a comprehensive survey on fundamental scene modeling tasks. 

\section{Scope of the report}
\label{sec:scope}
This report deals with 3D indoor scenes, which has a rich literature on different aspects of analysis techniques, and a relatively smaller literature on synthesis techniques. 
As such, it is hardly possible to exhaustively survey all such publications.
\revision{
This report is focused on providing technical insights into some of the prominent works in scene analysis and synthesis tasks, with an emphasis on how different scene representations necessitate the development of deep learning models that cater to such representation while addressing the scene modeling task at hand. 

Individual scene modeling tasks presented in this report deserve a survey of their own. Our aim is to provide directional pointers, with fundamental technical insights, on some of the seminal, popular, and recent works in these areas. To the best of our knowledge, this is the first such attempt to bring all scene modeling tasks in a single report, with a focus on neural network-based modeling (prominent model-driven approaches have also been touched upon where the context so necessitates).
}

%
%


\textbf{Organization} We first present different types of indoor scene representations (Section \ref{sec:layout_reps}), followed by various indoor scene datasets publicly available for use (Section \ref{sec:dataset_list}) for different analysis and synthesis techniques. 

In general, analysis of layouts, both 2D and 3D, spans a wide range of goals, from low-level understanding tasks such as primitive detection (corners, line segments) and semantic segmentation, to high-level layout understanding tasks such as saliency detection, layout reflowing, and layout retrieval to name a few. \newline 
In the context of 3D indoor scenes, analysis refers to understanding the object layout within a confined space, which can be categorized into two fundamental tasks -- 3D object detection (Section \ref{subsec:obj_det}) and 3D scene segmentation (Section \ref{subsec:3d_ssg}). A more high-level but challenging task in scene analysis we cover in this report is that of scene reconstruction, either from a single image or posed images (Section \ref{subsec:scene_recon}).  Finally, we discuss relevant literature in 3D scene similarity (Section \ref{subsec:scene_sim}), thus concluding our coverage of scene analysis tasks. For synthesis techniques (Section \ref{sec:synthesis}), we mainly look at recent progress towards this goal, which by default, has been skewed towards neural models. A more detailed discussion of model-driven techniques for scene generation can be found in \cite{chaudhuri2020learning}.

\textbf{Audience} This report is written keeping in mind new graduate students in computer science (and allied disciplines). Readers should have a basic understanding of linear algebra, probability and statistics, machine learning and standard deep learning machinery (ex: CNN, GCN, GAN, VAE). This report, is by no means, an exhaustive collection of works dealing in the analysis and/or synthesis of indoor scenes. It is more of a directional digest for research in this area, exposing the main problems and challenges involved, the observed gains due to a paradigm shift from model-driven approaches to data-driven ones where applicable, the interplay between scene representation and choice of computing machinery (neural network), and the evolving trends that could inspire novel problems in this area. Finally, we discuss open problems in this domain that have wider industrial applicability.

\section{Scene Representations}
\label{sec:layout_reps}
\begin{figure}[!t]
    \centering
    \includegraphics[width = \linewidth, height =  0.25\linewidth]{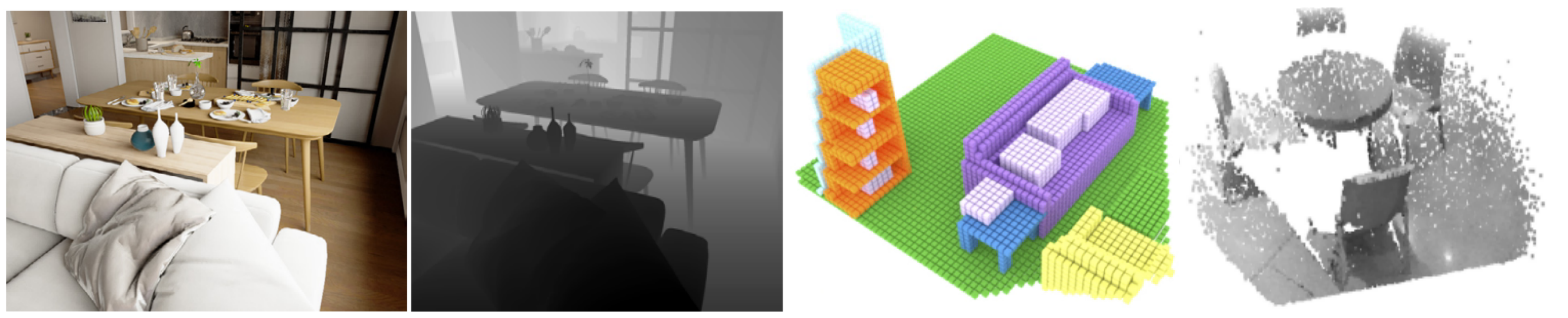}
    \caption{Different forms of visual representation for indoor scenes as discussed in Section \ref{subsec:visual_rep}: (First two) RGB image depicting an indoor scene and its corresponding depth map \cite{alhashim2018high}, (third) voxelized representation of an indoor scene \cite{song2017semantic} and (right) point cloud representing a subscene \cite{qi2020imvotenet}.}
    \label{fig:visual_reps}
    \vspace{-15pt}
\end{figure}
Representation of a 
scene should convey information about the \emph{combination} of atleast two things -- (1) the composition of its layout, either as a single entity or as a set of constituent objects (semantics) and, (2) the arrangement of objects (and perhaps their relations) in a given space. Such a combination could be either explicitly encoded or may need to be inferred separately.
Representations that necessitate additional processing to infer information about the object semantics and their placement are oblivious to the underlying structure of the layout, and are said to be purely visual in nature (ex: raster images in 2D, point clouds, and voxel grids in 3D). On the other hand, if the semantics of the constituent 
objects and their placement (and even relations) is explicitly encoded, such representations are said to be structural in nature (ex: multi-channel segmented images, graphs). Thus, we broadly classify 
scene representation into two categories: (a) visual, and (b) structural.

\subsection{Visual Representations}
\label{subsec:visual_rep}
Visual representations, as the name suggests, mainly represent a scene 
as a single entity. Common examples of this kind of representation include 2D images (monochrome, RGB and RGB-D) and 3D point clouds. Figure \ref{fig:visual_reps} shows these common visual representations used for scenes, and layout data, in general.

\textbf{2D images and 3D voxels} Two-dimensional rectangular grids of picture elements in the form of images are the common form of visual representation for scenes. 
Such data could be obtained in different ways such as using digital cameras or scanning devices (RGB-D cameras) such as Microsoft Kinect. There also exists a 3D counterpart to 2D pixel grids called 3D voxel grids that approximate a 3D surface. Such volumetric data representations are memory intensive and have been found to be intractable for modeling 3D shapes, let alone 3D scenes.

\textbf{3D point cloud} Real-world 3D indoor scenes are also digitized using commercial 3D scanners, where the acquired data is stored as a point cloud representing the surface of objects in the 3D environment, as shown in Figure \ref{fig:visual_reps}. Point clouds do not encode topological information of the underlying 3D content and simply depict the 3D data in the simplest visual form possible. 

This kind of 2D grid representation in the form of images and 3D point cloud respectively, depict an indoor scene as a single entity, i.e., the semantics of the constituent objects, their geometric arrangement and their relationships are not accounted for by the representation, and will have to be inferred separately. Examples of such representations include monochrome images for 2D indoor scenes in the form of a floorplan \cite{kalervo2019cubicasa5k}, RGB (+D) images \cite{Silberman:ECCV12} and point clouds \cite{dai2017scannet}. Figure \ref{fig:visual_reps} illustrates different visual representations of indoor scenes.

The above representations convey information about indoor scenes in a structure-agnostic manner -- that is, only an abstraction of the scene layout is available. Its composition based on constituent objects will need to be inferred separately. 
\subsection{Structural representations}
\label{subsec:struc_rep}
Structure refers to the atomic composition of an entity/matter. In the case of indoor scenes, it has to do with the type, arrangement, and/or relationship of different objects forming the scene layout. There are many ways of representing structured data, which have been surveyed \cite{chaudhuri2020learning} for 3D structures in general. While most of it directly applies to 3D scenes, the choice of such structural representations for different works chosen for this report needs to be discussed. We fill this piece of information below by briefly describing structure representations and the associated works in the context of 3D indoor scenes.
\begin{figure}[!t]
    \centering
    \includegraphics[width = \linewidth]{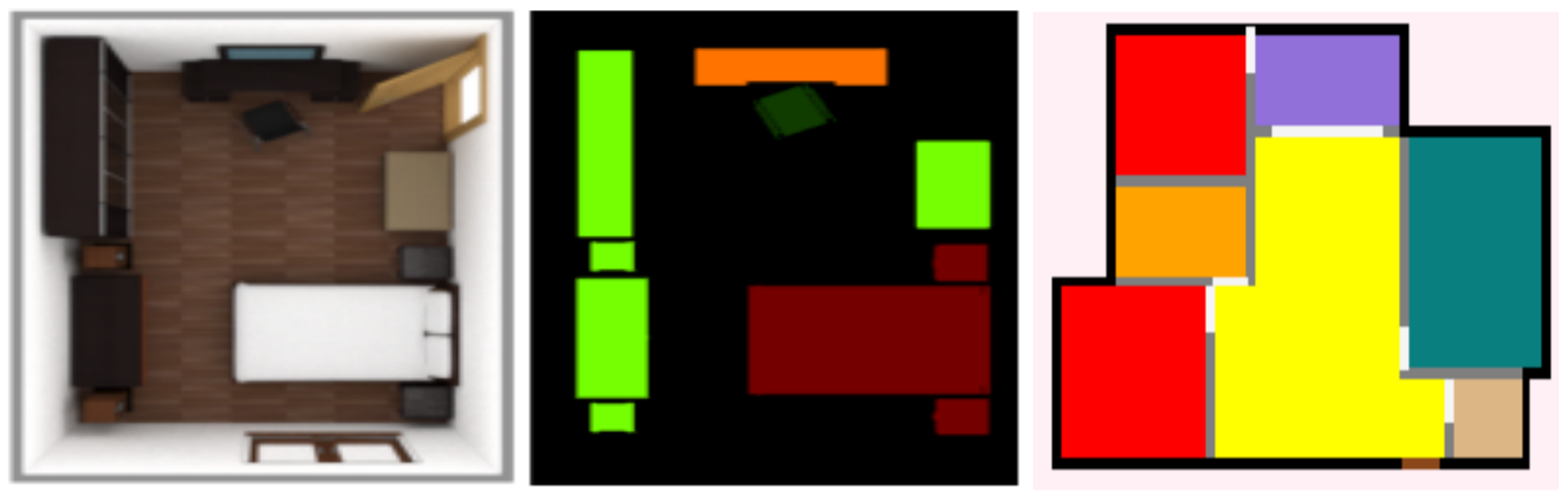}
    \caption{Figure illustrating semantic segmentation of indoor scenes:  a top-down scene image on the left \cite{wang2018deep} and a 2D floorplan on right \cite{wu2019data, patil2021layoutgmn}. Such segmented semantic entities form the simplest form of structural representation as discussed in \ref{subsec:struc_rep}.}
    \label{fig:segmented_geo}
    \vspace{-15pt}
\end{figure}

\textbf{Segmented Scenes}
Structure, in its simplest and weakest form, can be thought of as assigning semantic labels to visual representations of 2D, 3D scenes (images, voxels, point cloud etc.), where each pixel/voxel/point represents the type of room/object entity present at that location.
For example, assigning labels to (a) rooms in a floorplan image as in the RPLAN dataset \cite{wu2019data, patil2020read} and (b) objects (bed, nightstand, table lamp, cabinet etc.) in a top-down scene image \cite{wang2018deep} 
can be considered as collections of simplest scene structures. Figure \ref{fig:segmented_geo} shows a few examples of this form of structural representation.
This representation is built upon standard visual representations. Furthermore, machine learning models used for processing (analyzing and synthesizing) visual representations can be directly employed for these segmented layouts, making them easier to work with. The disadvantage is that they do not (and can not) understand the underlying relationships among different elements, and therefore, are not a strong fit for geometric analysis and synthesis tasks.

\textbf{Component/Entity Sets}
A set of indoor scene elements, i.e., objects, with information about their semantics and oriented bounding box, explicitly accounts for ``atoms'' in the structural representation. Such a set is called an entity set, which essentially is a set of freely-floating elements in space, with no relationship information encoded between any pair of elements. This kind of scene representation is used in conjunction with a wide range of neural networks that process just the elements (their box coordinates and semantics), as evidenced in sequence-to-sequence analysis techniques \cite{aggarwal2020form2seq} which makes use of a Recurrent Neural Network, or, for synthesis tasks such as in \cite{wang2020sceneformer, paschalidou2021atiss} which make use of a Transformer.

\textbf{Graphs}
Adding relationships between pairs of elements (nodes of a graph) in the entity set, in the form of edges, reveals the full structure of a scene layout \cite{patil2021layoutgmn, zhang2021holistic}. 
%
%
\begin{figure*}[!t]
    \centering
    \includegraphics[width = \linewidth, height = 0.275\linewidth]{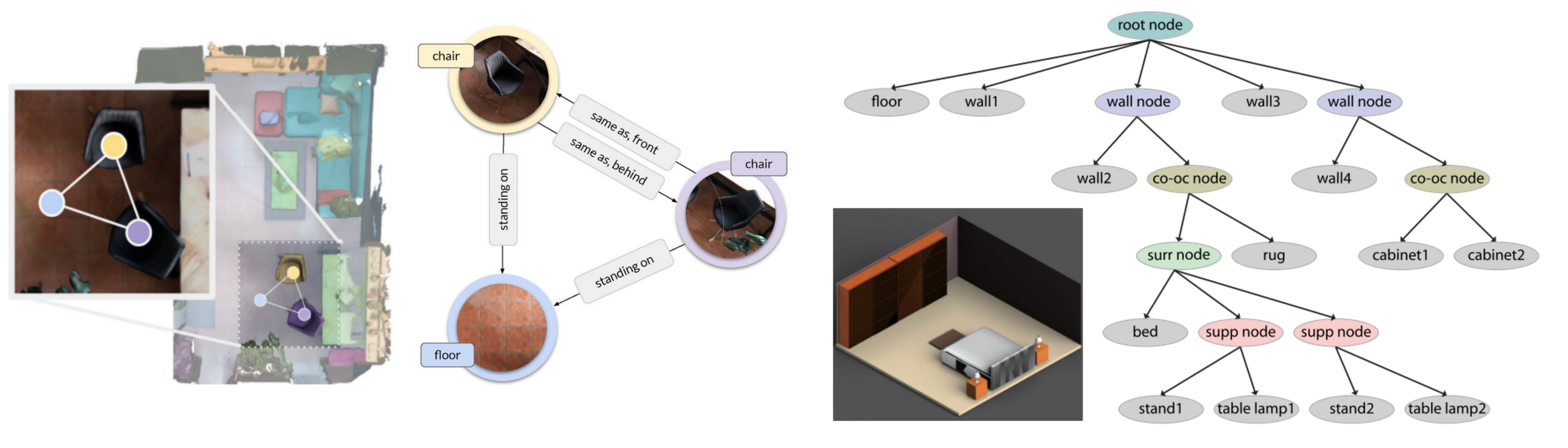}
    \caption{Representing 3D indoor scenes via strong structural representations (Section \ref{subsec:struc_rep}): on the left is an indoor scene represented as a semantic-relational graph \cite{wald2020learning}, and on the right is a bedroom scene represented as a hierarchy \cite{li2019grains}.}
    \label{fig:graph_rep}
\end{figure*}
These edges connecting nodes in a graph usually encode spatial relationships, such as adjacency, proximity \cite{lee2020neural, hu2020graph2plan} and physical support \cite{fisher2011characterizing}, but can also be simply connected between any two pairs of objects/elements, regardless of their spatial relationships \cite{manandhar2020learning, patil2021layoutgmn}. 
Figure \ref{fig:graph_rep} shows one such example of a semantic relationship graph for an indoor 3D scene.

The advantage of using graphs is that they are a more general and flexible form of structured representation. 
However, structured scene modeling is constrained by advances in graph modeling techniques, which is an active area of research in the broader machine learning community. As such, the development of sophisticated architectures for analysis tasks \cite{li2019graph} and generative models of arbitrary graphs is still in nascent stages.\\
%

Any flat graph can be encoded as a hierarchy by repeatedly contracting its edges. Trees or hierarchies, are therefore, less dense. The key difference is that hierarchies consist of internal nodes that represent groups of objects, while all the nodes of a graph represent the objects. Hierarchies are a class of restricted graphs and can be used to represent much of the naturally occurring structure in the real world. 
For example, a 3D scene can be thought of as a hierarchy of objects \cite{liu2014creating, li2019grains}, where objects are grouped based on their spatial positions, which in turn, is based on the functionality of individual objects in a group, see Figure \ref{fig:graph_rep}. This representation was also extended to 2D documents in \cite{patil2020read} where individual document entities were merged along a tree using spatial relationships.
A major bottleneck of representing 3D scenes 
using hierarchies is that there is no unique way of doing so, and as such, task-specific models that consume hierarchies inherit design limitations as a result of hard-coded heuristics used to construct such hierarchies.\\

\section{Indoor 3D scene datasets}
\label{sec:dataset_list}
There exist indoor scene datasets that either capture real-world scenes using acquisition devices or are professionally designed using curated 3D CAD models of furniture assets. Table~\ref{tab:dataset_list} summarizes various such indoor 3D scene datasets, which have been discussed in literature at various points such as in \cite{pintore2020state, li2021openrooms, fu20213d}. We aim to provide a comprehensive list of such up-to-date datasets along with the potential applications they could serve, each of which is briefly discussed below.

\begin{table*}[!t]
\centering
\resizebox{\textwidth}{!}{%
\begin{tabular*}{1.16\linewidth}{l|c|c|c|c|c|c|c}
\toprule
\textbf{Name} & \textbf{Data} & \textbf{Coverage} & \textbf{Capture} & \textbf{\#scenes} & \textbf{\#CAD models} & \textbf{Model Textures} & \textbf{3D Annotation} \\

    \midrule
    \multicolumn{8}{l}{\textbf{Real scans}}\\
    \midrule
    SUN 3D \cite{xiao2013sun3d}  & Registered RGB-D &  Perspective & Hand-held video & $254$ & - & No texture & Raw PCD \\

    SUN RGB-D \cite{song2014sliding}  & Registered RGB-D &  Perspective & Hand-held video & $10779$ & - & No texture & Raw PCD \\
    Matterport3D \cite{chang2017matterport3d} & Registered RGB-D &  Panoramic & Tripod & $2,056$ & - & Rec. from scans & Raw Mesh\\
    ScanNet \cite{dai2017scannet} & Registered RGB-D &  Perspective & Hand-held video & $1,506$ & $296$ & Rec. from scans & Raw Mesh \\
    Scan2CAD \cite{avetisyan2019scan2cad} & Mesh &  All & Manual modeling & 1506 & 3049 & No texture & Mesh \\

    OpenRoom \cite{li2021openrooms} & RGB-D &  Perspective & Hand-held decide & 1287 & 44 & UV mapping & Mesh \\
    %
    UZH 3D \cite{UZHdataset} & Registered PCD & Scan & Tripod & 53 & - & No texture & PCD\\
    3DSSG \cite{wald2020learning} & PCD & All & Hand-held device & 1482 & - & Rec from scans & Mesh + scene graph\\

    SceneNN \cite{hua2016scenenn} & RGB-D & Perspective & Hand-held device & 100 & - & Rec. from scans & Mesh\\
    \midrule
    \multicolumn{8}{l}{\textbf{Synthetic}}\\
    \midrule
    Replica \cite{replica19arxiv} & CAD model &  All & Manual modeling & 18 & - & RGB texture camera & Mesh\\

    Structured3D \cite{zheng2020structured3d} & CAD model &  All & Manual modeling & 3500 & - & No texture & 3D structure\\

    SceneNet \cite{handa2016understanding} & Mesh &  All & Manual modeling & 57 & 3699 & No texture & Mesh \\

    InteriorNet \cite{li2018interiornet} & RGB &  Perspective & Manual modeling & - & - & No texture & - \\

    Hypersim \cite{roberts2021hypersim} & RGB &  Perspective & Manual modeling & 461 & - & Per pixel color & RGB-D \\

    3D-FRONT \cite{fu20213d} & Mesh &  All & Manual modeling & 18968 & 13151 & Professional & Mesh\\
    \midrule
    \multicolumn{8}{l}{\textbf{Real scene images}}\\
    \midrule
    ETH 3D \cite{ETH3D} & Registered RGB &  Perspective & Tripod & 898 & - & No Texture & PCD\\

    CRS4/ViC \cite{CRS4} & Registered RGB &  Panoromic & Tripod & 191 & - & No texture & - \\
    NYU Depth v2 \cite{silberman2012indoor} & Registered RGB-D &  Perspective & Hand-held video & 1449 & - & No texture & RGB-D \\
    TUM \cite{sturm2012benchmark} & Registered RGB-D &  Perspective & Hand held video & 39 & - & No texture & RGB-D \\

    \bottomrule
\end{tabular*}%
}
\caption{
A summary of publicly available 3D indoor scenes datasets, grouped based on acquisition source, along different axes that include high-level details such as the physical mode of capture to low-level ones such as the kinds of annotations on the scene and the number of CAD models/scenes. \textit{\#scenes} indicates number of rooms/scenes populated with 3D furniture objects, \textit{PCD}=``Point cloud".
}
\label{tab:dataset_list}
\vspace{-15pt}
\end{table*}

%

\textbf{SUN 3D}~\cite{xiao2013sun3d} offers a dataset of large-scale RGB-D video frames with semantic object segmentation and camera pose. The dataset contains 415 videos captured for 254 different indoor spaces, in 41 different buildings. Geographically, the places scanned are mainly distributed across North America, Europe, and Asia. The dataset can be used to obtain (a) a point cloud of the scene; (b) 3D object models obtained from segmentation; (c) all viewpoints of an object, and corresponding camera poses relative to that object; (d) a map of a room, showing all of the objects and their semantic labels from a bird’s-eye view.

\textbf{UZH 3D dataset}~\cite{UZHdataset} contains 40 laser-scanned models of office environments and apartments. Some scenes have arbitrarily oriented walls which pose a challenge to many techniques in reconstructing floor plans. The point cloud models in the dataset are provided in ASCII PTX format with color information.

\textbf{ETH 3D dataset}~\cite{ETH3D} consists of 898 RGB images of both indoor and outdoor spaces. The dataset also provides ground truth point cloud and depth maps which can be used to benchmark multi-view stereo algorithms.

\textbf{Matterport3D}~\cite{chang2017matterport3d} provides a large-scale RGB-D dataset of 90 building-scale scenes. The dataset contains 10,800 panoramas and 194,400 RGB-D images. It is also provided with reconstructed textured 3D mesh with object-level semantic annotations and camera pose. The dataset can be used for various tasks such as room-type classification, semantic segmentation, surface normal estimation, keypoint matching, and view overlap prediction.

\textbf{ScanNet}~\cite{dai2017scannet} is a RGB-D video dataset of 1513 indoor scenes. The 3D reconstructed mesh has texture information and is labeled with object-level semantic segmentations. Moreover, the dataset also provides aligned 3D CAD models for a subset of scans. The dataset can be used for many 3D scene understanding tasks including 3D object classification, semantic voxel labeling, and CAD model alignment and retrieval.


\textbf{CRS4/ViC dataset}~\cite{CRS4} contains equirectangular RGB images 
covering 360x180 degrees of multi-room residential as well as commercial environments. It also contains images of rooms with double-sloped ceilings and cluttered with many objects, making it a challenging dataset to use for reconstructing a 3D floor plan. 

\textbf{Replica dataset}~\cite{replica19arxiv} provides 3D indoor scene reconstructions of rooms and buildings with a rich semantic variety of environments and their scale. The dataset contains high-dynamic-range (HDR) textures and per-primitive semantic class and instance information. Due to the high level of realism of renderings from the Replica dataset, the creators believe that deep learning models trained on this dataset can adapt well to real-world images and videos of indoor scenes.

\textbf{Structured3D dataset}~\cite{zheng2020structured3d} contains rich ground truth 3D structure annotations of 21,835 rooms in 3,500 houses, and more than 196k photo-realistic 2D renderings of the rooms. The scenes are represented in the format of ``primitive + relationship". The usefulness of the dataset is demonstrated on room layout estimation task.

\textbf{NYU Depth v2 dataset}~\cite{silberman2012indoor} consists of 1449 RGB-D images of commercial and residential buildings comprising of 464 indoor scenes. The dataset provides dense per-pixel labeling, where each object in the image is labeled with class label and instance annotations. The dataset also includes \textit{support} annotations between two objects in the image. This dataset can be used for tasks such as object recognition, segmentation, and inference of physical support relationships.

\textbf{SUN RGB-D dataset}~\cite{song2014sliding} consists of 10779 RGB-D images of real indoor scenes captured using four different sensors. The entire dataset is densely annotated with room category, 2D and 3D oriented bounding boxes for objects, and camera pose information. Specifically, it includes 146,617 2D polygons and 58,657 3D bounding boxes with accurate object orientations, as well as a 3D room layout and category for scenes. This dataset can be used for scene-understanding tasks and evaluate such models using meaningful 3D metrics.

\textbf{TUM dataset}~\cite{sturm2012benchmark} contains 39 image sequences capturing office environments and industrial halls. Each sequence contains color and depth images and also ground truth trajectory. The dataset is aimed at evaluating visual odometry and visual SLAM systems.

\textbf{3DSSG dataset}~\cite{wald2020learning} provides 3D semantic scene graphs for 1482 scenes from 3RScan~\cite{Wald2019RIO} dataset. 3DSSG dataset contains scene graphs with 40 different types of object relationships, and 93 different attributes for objects from 534 different class labels represented in class hierarchies. Such semantically rich scene graphs can be used for many applications such as semantic scene graph prediction and cross-domain scene retrieval tasks.

\textbf{SceneNN}~\cite{hua2016scenenn} is a dataset of RGB-D scans of 100 indoor spaces. The dataset provides information about camera pose, reconstructed mesh, color and texture information, axis-aligned and oriented bounding boxes, as well as object pose. This dataset can be used for shape completion, scene relighting, creating synthetic scenes using CAD models by using object distribution statistics of real scenes from the SceneNN dataset, and novel view synthesis tasks.

\textbf{Scan2CAD}~\cite{avetisyan2019scan2cad} is a large-scale dataset of 1506 ScanNet~\cite{dai2017scannet} scene objects aligned to 14225 (3049 unique) CAD models of ShapeNet dataset~\cite{chang2015shapenet}. It contains 97607 pairwise keypoint correspondences between scene objects and CAD models. The dataset also contains oriented bounding boxes for objects in the scenes. This information can be used in various applications such as correspondence prediction between unseen 3D scenes and CAD models, and their pose estimation task.

\textbf{OpenRoom}~\cite{li2021openrooms} dataset is aimed at creating photo-realistic indoor scenes by adding high-quality material and lighting information. The dataset uses 1287 ScanNet~\cite{dai2017scannet} scenes to create such photo-realistic indoor scenes. The dataset is annotated with ground truth scene layout, high-quality material, and spatially-varying BRDF lighting, including direct and indirect illumination, light sources, per-pixel environment maps and visibility. This dataset is useful in inverse rendering, scene understanding, and robotics applications. The dataset can also be used for shape, material and lighting estimation which are crucial in augmented reality (AR) and virtual reality (VR) applications. 

\textbf{SceneNet}~\cite{handa2016understanding} is a dataset of synthetically generated 3D indoor scenes. It contains 57 scenes of five categories: bedroom, office, kitchen, living room, and bathroom. Each scene has 15-250 objects. The RGB-D renderings of these scenes can be used for per-pixel semantic segmentation tasks. The dataset also provides a tool that can be used to generate unlimited labeled 3d indoor scenes programmatically, which is helpful in training data-driven machine learning models.

\textbf{InteriorNet}~\cite{li2018interiornet} is a synthetic 3D indoor scene dataset created using 1M furniture CAD models and 22M interior layouts. The dataset has 15K sequences of 10K randomly selected layouts and 5M images rendered from 1.7M layouts. The dataset can be used to train and evaluate SLAM systems. 

\textbf{Hypersim}~\cite{roberts2021hypersim} is photo-realistic synthetic 3D indoor scene dataset. It contains 77400 images rendered from 461 indoor scenes with per-pixel labels, ground truth scene geometry, material and lighting information, and semantic segmentation label. The dataset was evaluated on two scene understanding tasks: semantic segmentation and 3D shape prediction.

\textbf{3D-FRONT}~\cite{fu20213d} dataset contains professionally designed 3D indoor scenes of 31 scene categories. It has 6813 CAD houses with 18968 rooms furnished with high-quality textured 3D models from 3D-FUTURE~\cite{fu20213dfuture} dataset. The usefulness of the dataset was demonstrated on scene understanding tasks such as 3D indoor scene synthesis and object texturing in scene context.

\revision{
The explosion of NeRF \cite{mildenhall2021nerf} has brought a variety of scene images into focus, most of which are single-object images and are not catered to indoor scenes. Here we briefly touch upon a few datasets used in novel view synthesis. 

\textbf{RealEstate10K}~\cite{zhou2018stereo} is a dataset of camera poses on 10K real estate YouTube videos that contain indoor and outdoor scenes of houses. These videos are divided into clips of 1-10 seconds, and for each clip, the dataset provides information such as camera position, orientation, and field of view per frame. This dataset finds its usefulness in tasks related to view synthesis.

\textbf{ACID}~\cite{liu2021infinite} Another commonly used dataset for view synthesis is the Aerial Coastline Imagery Dataset (ACID)~\cite{liu2021infinite}. It is a dataset of outdoor nature videos (891 videos) annotated with camera pose information.

\textbf{Common Objects in 3D (Co3D)}~\cite{reizenstein21co3d} is another dataset consisting of 18,619 videos of objects from 50 MS-COCO categories. Compared to RealEstate10K and ACID, the Co3D dataset is simpler as the videos are focused on single objects with no occlusion. 

\textbf{Ego4D}~\cite{grauman2022ego4d} provides a bit different dataset with videos capturing everyday activities from first-person perspective. It consists of 3025 hours of videos shot in indoor and outdoor scenarios. In addition to videos, it also provides other information such as 3D scans, audio, gaze, stereo, multiple synchronized wearable cameras, and textual narrations. This dataset finds usefulness not only in view synthesis but also in other challenging tasks such as analyzing hand-object interaction, audio-visual conversation, and forecasting activities.
}
\section{3D scene analysis}
\label{sec:analysis}

\begin{figure}[!t]
    \centering
    \includegraphics[width=0.85\linewidth, height=0.7\linewidth]{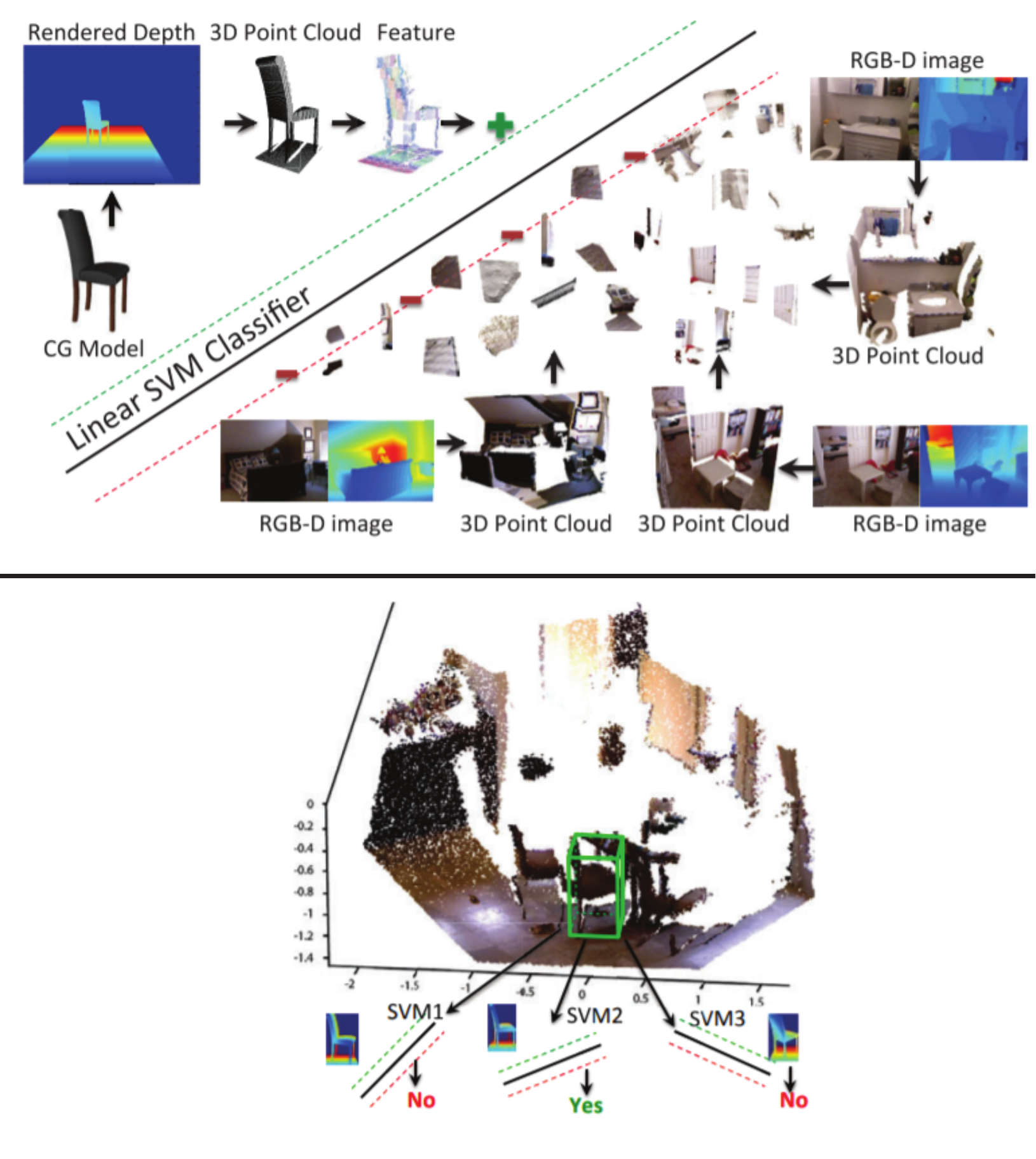}
    \caption{Illustration of training (top) and test (bottom) phases for 3D object detection using \emph{Sliding Shapes} \cite{song2014sliding}. Refer to the text in Section \ref{subsec:obj_det} for details.}
    \label{fig:sliding_shapes}
    \vspace{-15pt}
\end{figure}


%
\begin{table*}[!t]
\resizebox{\textwidth}{!}{%
\begin{tabular*}{1.28\linewidth}{l|c|c|c|c|c|c|c}
\toprule
\textbf{Related Work} & \textbf{\makecell{Learning\\ framework}} & \textbf{Scene rep} & \textbf{Backbone} & \textbf{Input} & \textbf{Output} & \textbf{Dataset} & \textbf{Evaluation Metric(s)}\\[0.5 ex]
\midrule
%
\cite{song2014sliding} & Supervised & RGB-D image & SVM & \makecell{TSDF + 3D Normal \\+ Point Density \\+ Shape features} & $Pr$(C) for a 3D OBB & SUN RGBD & AP; 2D,3D IoU \\
%
\midrule
\cite{song2016deep} & Supervised & RGB-D image & 3D CNN & \makecell{TSDF scene \\and RGB image} & $Pr$(C) and 3D OBB & SUN RGBD & AP, AR, 3D IoU\\
%
\midrule
\cite{ren2016three} & Supervised & RGB-D image & SVM & \makecell{Point density \\and normal features} & 3D OBB & SUN RGBD & AP, AR\\
%
\midrule
\cite{qi2019deep} & Supervised & RGB-D image & PointNet++ & 3D point cloud & 3D OBB &  SUN RGBD, ScanNet & 3D IoU, AR, mAP \\
%
\cite{qi2020imvotenet} & Supervised & RGB-D image & PointNet++, 2D CNN & \makecell{RGB image and\\ 3D point cloud} & 3D OBB & SUN RGBD & 3D IoU, AP \\
%
\midrule
\cite{xie2020mlcvnet} & Supervised & RGB-D image & PointNet++ & 3D point cloud & 3D OBB & SUN RGBD, ScanNet & 3D IoU, mAP\\
%
\midrule
\cite{zhang2020h3dnet} & Supervised & RGB-D image & PointNet & 3D point cloud & 3D OBB & SUN RGBD, ScanNet & 3D IoU, mAP\\
%
\midrule
\cite{liu2021group} & Supervised & RGB-D image & PointNet, Transformer & 3D point cloud & 3D OBB & SUN RGBD, ScanNet & 3D IoU, mAP \\
%
\midrule
\cite{yu2022rotationally} & Supervised & RGB-D image & \makecell{PointNet and \\Equivariant Point Network} & 3D point cloud & 3D OBB & Sun RGBD, ScanNet & 3D IoU, AP, mAP \\
\bottomrule
%
%
\end{tabular*}%
}
\caption{Table summarizing prominent \textbf{3D object detection works} in indoor scenes. We provide details on the scene representation, the input for and output of the model, the central algorithm that makes the task possible, dataset used and the metrics employed to evaluate results from proposed methods. AP - Average Precision, mAP - mean of Average precision, AR - Average Recall, IoU - Intersection over Union.} 
\label{tab:3d_object_detection_works}
\vspace{-15pt}
\end{table*}

The first step in computational scene synthesis, i.e., teaching computers to generate indoor environments, is scene analysis, i.e., teaching computers to understand their composition -- what characterizes scenes of a particular category (say, bedrooms), what kind of furniture goes in there, how to identify different furniture objects, how to detect different instances of the same object in the scene, and how to reason about their placements in the context of global scene plausibility. In other words, computational analysis of scenes, a.k.a scene understanding, is a prelude to scene synthesis. 

Understanding comes from observations, which are relayed by indoor scene datasets, which have been discussed in Section \ref{sec:dataset_list}. 
Each of these datasets uses a different form of representation for indoor scenes, as categorized in Section \ref{sec:layout_reps}. In literature, different scene analysis tasks use different kinds of representation, which could be motivated by different factors such as the easy availability of a dataset with one form of representation, friendliness toward off-the-shelf networks used as a part of the proposed approach, or the need for developing novel architectures due to the choice of a certain representation. In the upcoming sections, we provide a summary of different works along similar axes. \\
The main analysis tasks we cover in this report include 3D object detection (Section \ref{subsec:obj_det}), semantic scene segmentation (Section \ref{subsec:3d_ssg}), 3D scene reconstruction (Section \ref{subsec:scene_recon}, and 3D scene similarity (Section \ref{subsec:scene_sim}).

\begin{figure}[!t]
    \centering
    \includegraphics[width=0.85\linewidth, height=0.7\linewidth]{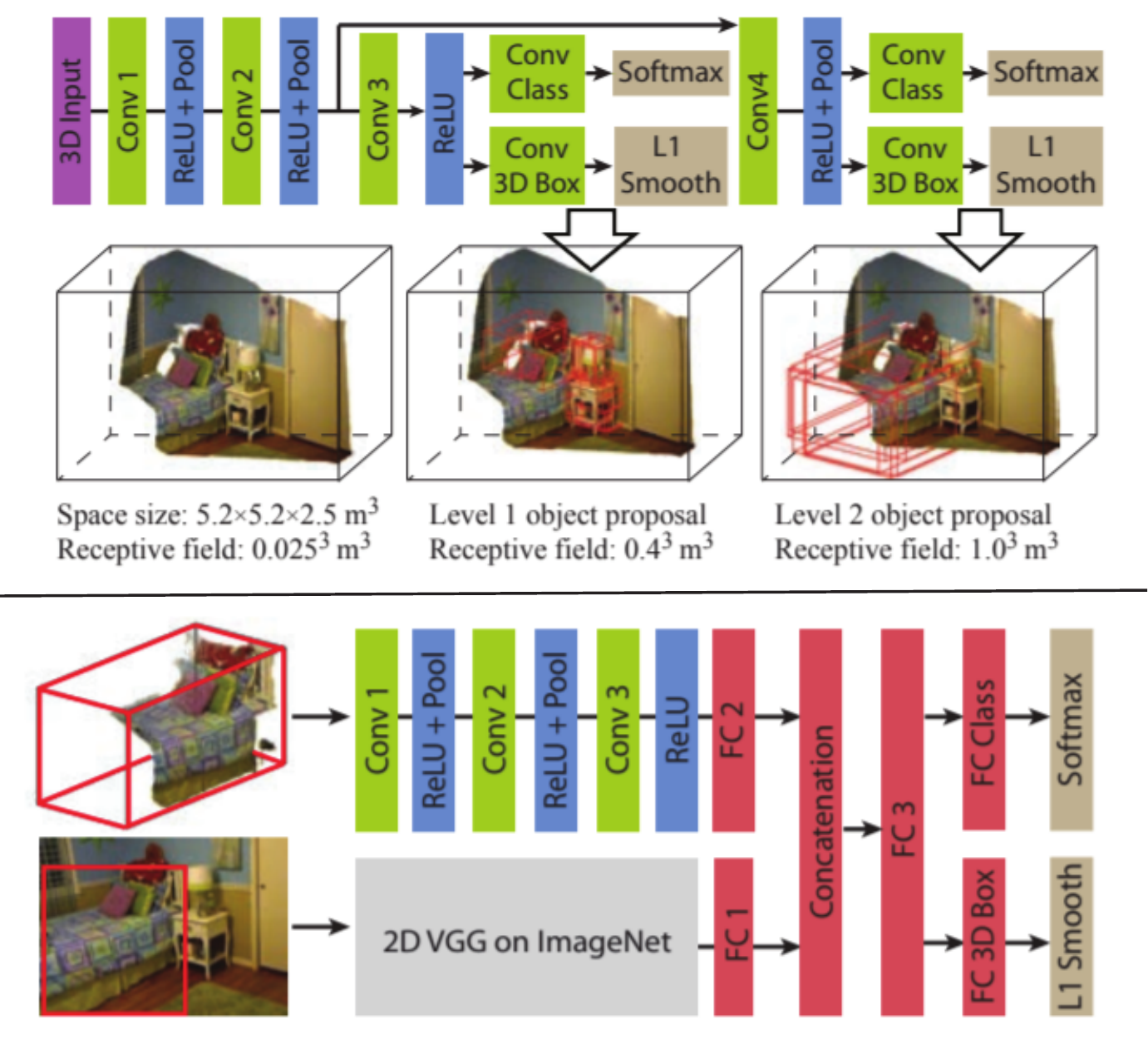}
    \caption{Figure illustrating the technique behind \emph{Deep Sliding Shapes} \cite{song2016deep}. Top: using a 3D CNN to propose regions of interest in 3D. Bottom: training pipeline based on the 3D and 2D CNNs to localize bounding boxes in 3D space. }
    \label{fig:deep_sliding_shapes}
    \vspace{-15pt}
\end{figure}

\subsection{3D object detection}
\label{subsec:obj_det}
Recognizing objects in a scene, i.e., identifying their semantic category and localizing their spatial position via 2D/3D bounding boxes has been a fundamental and long-standing goal of computer vision. This report does \emph{not} cover works on 2D object detection in RGB images. Rather, we focus on notable works on 3D object detection in indoor scenes, a task that is challenging mainly due to variations in shape (both inter and intra-class), texture, illumination, viewpoint, and the presence of clutter and occlusions. Table \ref{tab:3d_object_detection_works} provides a comprehensive overview of notable methods on 3D object detection in indoor scenes. Broadly, these works can be categorized into three types: sliding window techniques \cite{song2014sliding, song2016deep}, grouping techniques \cite{qi2019deep,qi2020imvotenet, xie2020mlcvnet}, and group-free techniques \cite{liu2021group}, as discussed below.

\textbf{Sliding window techniques }
Song et al. \cite{song2014sliding} introduce Sliding Shapes, a supervised machine learning-based approach for 3D object detection. The work makes use of depth maps for designing a 3D object detector, in addition to a collection of 3D CAD models, where each CAD model is rendered from many viewpoints, obtaining synthetic depth maps for every viewpoint. For each depth rendering of a CAD model, features based on truncated sign distance fields (TSDF) values, 3D normals, point density, and voxel occupancy are extracted (collectively known as point features) and an exemplar support vector machine (Exemplar-SVM) classifier is trained on point features of the sensor-acquired scenes. During test time, a sliding window is moved through the 3D scene space to detect an object; see Fig \ref{fig:sliding_shapes}. 

The successor to Sliding Shapes, termed, Deep Sliding Shapes, was presented in \cite{song2016deep}. It is a supervised deep learning-based framework which makes use of a 3D ConvNet that takes a 3D volumetric scene from an RGB-D image as input and outputs 3D object bounding boxes. A 3D Region Proposal Network (RPN) is trained at two different scales to learn object-ness from geometric shapes. An Object Recognition Network (ORN) is \emph{jointly} trained with RPN to extract geometric features in 3D and color features in 2D, to eventually output a category label and 3D box coordinates. Figure \ref{fig:deep_sliding_shapes} illustrates these two steps during training. At test time, a sliding window is again moved through the space of a 3D scene to detect the presence of an object.

One main limitation of both the above approaches is that they do not explicitly encode object orientation, which can hurt the performance of a 3D object detection system. Ren et al. \cite{ren2016three} overcome this limitation by designing a new set of features, called, cloud-of-oriented-gradient (COG), that robustly link 3D object pose to 2D image boundaries. COG features are nothing but the gradients of 2D projections of oriented cuboid points falling inside the object voxel. COG features, in addition to the point cloud density features and 3D normal histogram features form the point features, are used to train an SVM for 3D object detection similar in spirit to Sliding Shapes. Sedaghat et al. \cite{sedaghat2016orientation} also address the limitation of Deep Sliding Shapes by adding orientation classification as an auxiliary task, and demonstrate that speed and accuracy of 3D detection using a sliding window increases when the 3D CNN is jointly trained on object labels, location and pose.

\begin{figure}[!t]
    \centering
    \includegraphics[width = \linewidth, height = 0.6\linewidth]{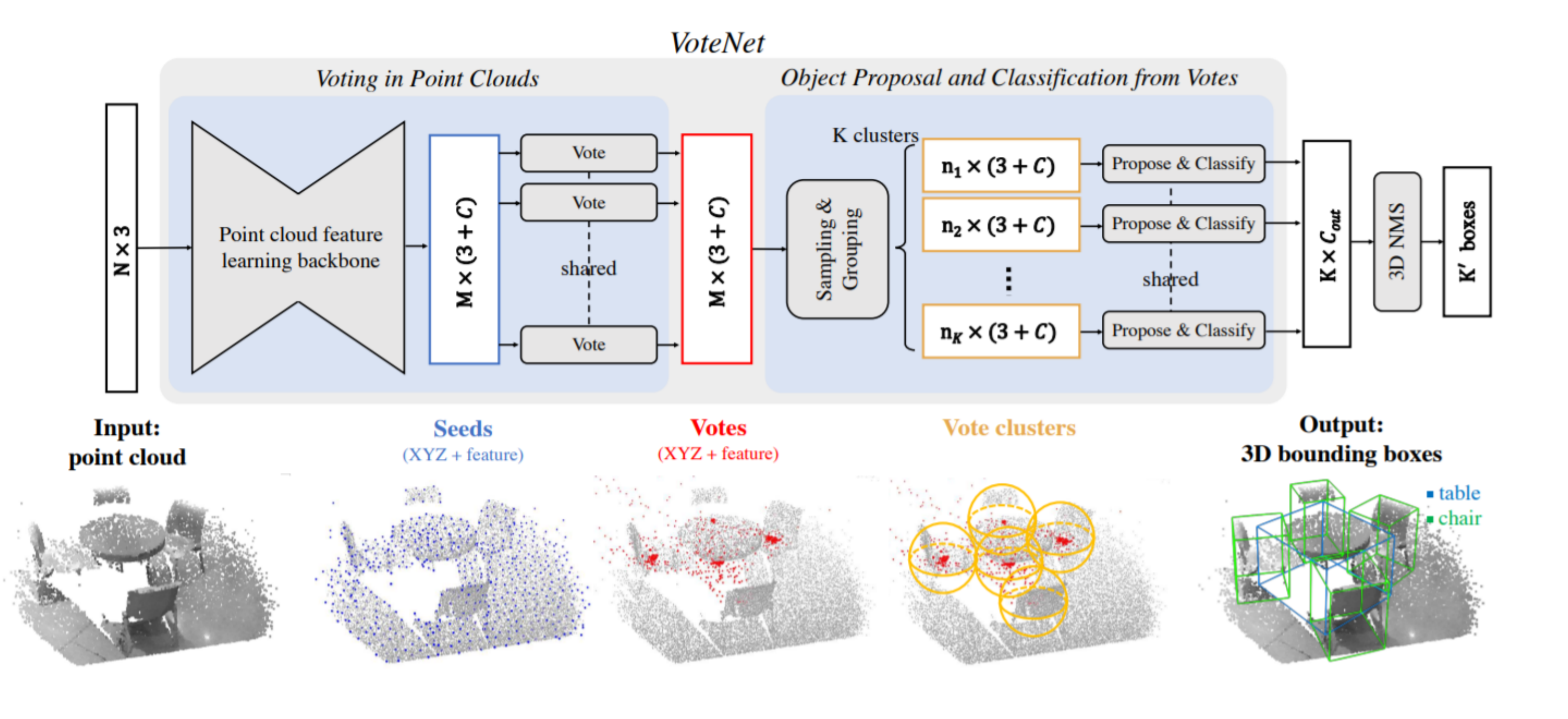}
    \caption{An end-to-end learning pipeline for 3D object detection using voting technique (\emph{VoteNet}) \cite{qi2019deep} that directly operates on 3D data without the need for any 2D image priors, such as 2D object detectors. }
    \label{fig:vote_net}
    \vspace{-15pt}
\end{figure}

\textbf{Voting techniques }
In recent years, voting concepts, specifically, Hough voting, have made a comeback in the space of 3D object detection. VoteNet \cite{qi2019deep} and ImVoteNet \cite{qi2020imvotenet} are two such works that are built on voting strategies. 
VoteNet demonstrates two advantages of using the voting strategy for 3D object detection -- first, it does not make use of any 2D object detectors which used to be the de-facto step in 3D object detection, and second, the Hough voting technique used in the work is now differentiably learned in an end-to-end supervised manner. 
As shown in Figure \ref{fig:vote_net}, the input to the system is a colorless point cloud of a scene, which is processed by PointNet++ \cite{qi2017pointnet++}, to produce features for every point. A voting net, which is nothing but a MLP on these point features, produces virtual points, called votes, for centers of 3D bounding boxes. The votes are clustered in the 3D space using farthest point sampling and L$_2$ distance, from where the extent of the 3D bounding boxes and their centroids are regressed using another MLP. All of this training is done in a supervised manner, where the votes on the training data are available since it is supervised. ImVoteNet \cite{qi2020imvotenet} incorporates all the steps from VoteNet, but in addition, makes use of a 2D object detector, where votes are obtained in the image space, which is lifted to the 3D space (along with the 2D object center) by ray-casting.

\textbf{Hybrid techniques }
A hybrid model for 3D object detection on colorless scene point clouds was proposed by Zhang et al. \cite{zhang2020h3dnet}, where bounding boxes are represented using three geometric primitives -- bounding box centers, face centers, and edge centers. These hybrid geometric primitives represent an overcomplete set of constraints that are predicted using a neural network. The predicted geometric primitives are converted into object proposals by defining a distance function between an object and the geometric primitives. The main purpose served by the distance function is that it helps in the continuous optimization of object proposals. A final matching and refinement module is proposed to classify object proposals into detected objects.

\begin{figure}[!t]
    \centering
    \includegraphics[width = \linewidth, height = 0.6\linewidth]{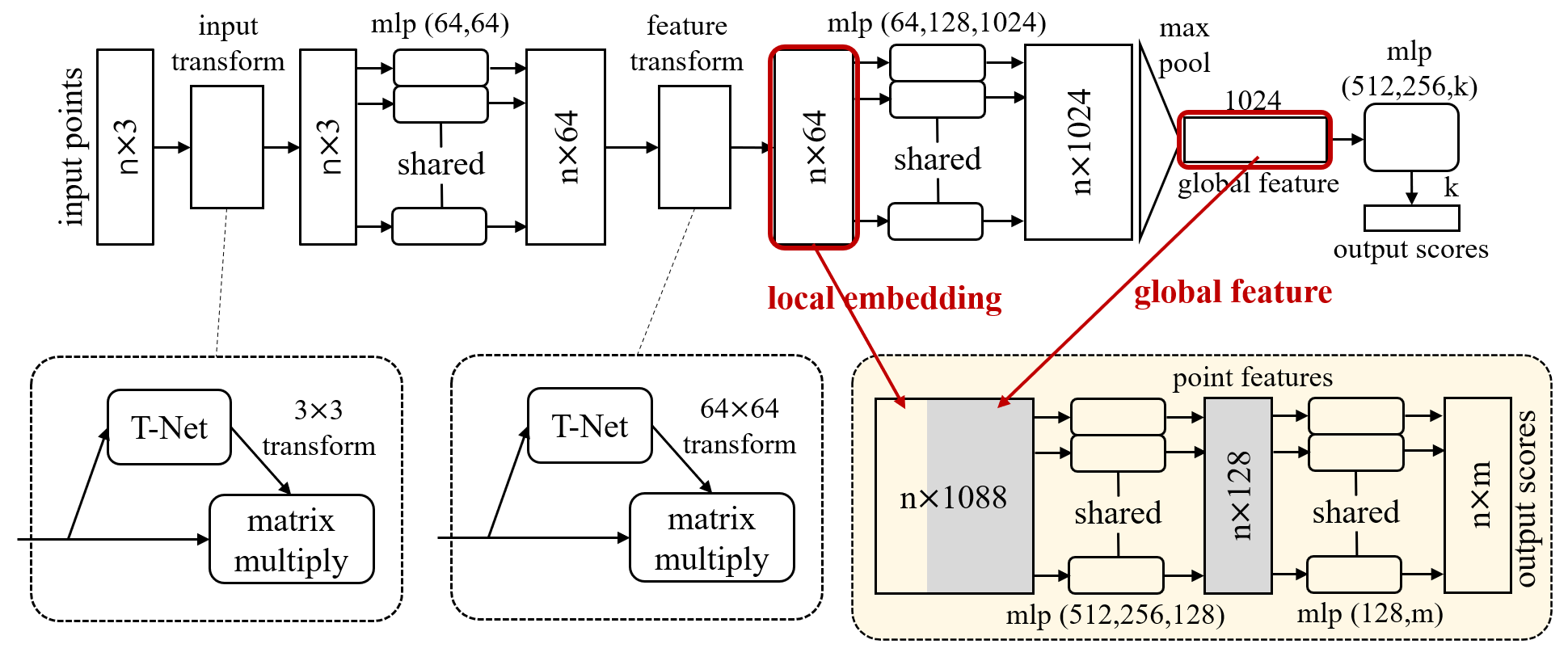}
    \caption{\emph{PointNet} architecture \cite{qi2017pointnet} that was originally proposed for the task of shape classification and segmentation at the object part level, as well as for indoor scenes. This work has been the basis for many 3D object detection works, especially for developing task-specific feature descriptors.}
    \label{fig:point_net}
    \vspace{-15pt}
\end{figure}

\revision{
\textbf{Context-aware techniques }
Despite a large inter- and intra-class shape variation and illumination effects in 3D indoor scenes, the context relation of object arrangements provides useful cues for object detection. Some prior works seek holistic scene understanding to improve object detection with other auxiliary tasks. For example, Lin et al.~\cite{LinFU13} first estimate the candidate 3D cuboids of the objects, and then use a conditional random field (CRF) model to \emph{jointly} solve for the scene classification and 3D object recognition. This holistic approach takes the scene context into account while also accounting for relations between different objects for the task of object classification. 

Zhang et al.~\cite{ZhangBKIX17} integrates the context relations into neural networks using automatically constructed scene templates. They first select a template and align it with the input scene by a transformation network, and then compute the global and local features of the input scene based on the aligned template. What makes this approach holistic is that it uses global features to classify scenes, while using both the global and local features to predict the existence of objects and the box offset for better alignment. Huang et al.~\cite{huang2018cooperative} refer to their approach of jointly recovering the object bounding boxes, room layout, and camera pose as a holistic one. They estimate all this information from a given RGB image, then use the predicted camera pose to project the inferred 3D bounding boxes back to the 2D plane, in order to obtain a more consistent prediction.

Recent investigations have looked into utilizing context information to complement 3D object detection. Feng et al.~\cite{FengGWZM21} take PointNet++~\cite{qi2017pointnet} as the backbone to generate candidate 3D object bounding boxes, and then use the object-object relation graph to reduce uncertainty during 3D bounding box regression. Duan et al.~\cite{Duan0LYL022} argue that 3D object detection can be improved by adopting relations between representative proposals, which is more efficient than those between all the predicted proposals. They accomplish this by proposing what they call a DisARM module, which first samples relation anchors with rich information and then estimates the weight of each proposal w.r.t the anchors based on spatial- and feature-aware displacements. The weighted proposal-anchor features provide contextual information to complement the anchor proposal feature. In addition, Sun et al.~\cite{0003FZLL22} propose an online data augmentation pipeline based on the functional relation between objects, called Correlation Field, that helps boost the performance of object detection.
}

\textbf{Transformer-based techniques }
A more recent supervised learning approach by Liu et al. \cite{liu2021group} to 3D object detection on scene point clouds makes use of the transformer model \cite{vaswani2017attention}, which is essentially an attention-based feature aggregation module on the input. Unlike voting-based object detection works such as VoteNet \cite{qi2019deep} and ImVoteNet \cite{qi2020imvotenet} where the points are assigned to an object candidate via a heuristic point-grouping stage, 
\cite{liu2021group} uses a grouping-free approach for detecting objects in a scene point cloud. Instead of obtaining a candidate object feature from a heuristically-grouped set of points (also called "votes"), candidate features are computed by neurally estimating the contribution of each point to the object candidate using attention mechanism. \\

Note that all the above works have been discussed in the context of indoor scenes, which is the focus of this report. There exist impactful works that have been developed in the context of outdoor scenes, specifically on the KITTI dataset \cite{geiger2013vision}, such as PointRCNN \cite{shi2019pointrcnn}, PointPillars \cite{lang2019pointpillars}, CenterPoint \cite{yin2021center} that operate on the 3D point cloud of the scene and are able to predict a 3D bounding box for scene objects. In theory, these architectures could very well be extended to 3D indoor scenes, but supporting experiments have not been presented.

\textbf{Discussion }
With an anticipated shift from feature-engineered methods \cite{song2014sliding, ren2016three, sedaghat2016orientation} to deeply learned ones \cite{song2016deep, qi2019deep, qi2020imvotenet, zhang2020h3dnet, liu2021group}, 3D object detection in indoor scenes has drawn significant research interest over the years, leading to improved performances of proposed approaches. 

\agp{ 
Assuming good quality indoor scene data at hand, extracting rich features from the 3D representation plays a critical role in the success of 3D object detection algorithms. Apart from the 3D-aware local shape features, the model should be designed to encode contextual scene information, either explicitly as done in sliding window techniques used in \cite{song2014sliding, song2016deep}, or encoding contextual relations as done in \cite{ZhangBKIX17, huang2018cooperative, FengGWZM21}, or implicitly as done in transformer-based networks. Encoding such contextual information proves useful when the input scene contains noise or occlusion and clutter. This means that the extracted features should account for spatial and semantic scene information to facilitate accurate object localization in the form of a detected 3D bounding box (location, size, and orientation). 

Since transformer models are well known for their ability to capture long-range dependencies via the self-attention mechanism,  
they can be employed on large-scale scenes without the loss of spatial information, as opposed to when using convolutional architectures with or without specially designed modules. This is evident from \cite{liu2021group}. The downside of transformer models is that the intermediate computational operations cannot be traced as easily as in the case of CNNs (ex. being able to visualize the intermediate layers and activation maps). There is an accuracy and interpretability trade-off with transformers, but given the improved performance, they are a good starting point.\\
\newline
} 
In addition, a strongly desired property in 3D deep learning is rotation equivariance. Accounting for equivariance to object rotations in 3D scenes, not at the global input level, but rather at the object level, is an interesting future direction -- attempts have been made to take into account object rotations when developing a 3D object detector \cite{ren2016three, sedaghat2016orientation}, but has not been explored by deeply learned methods. 

Since we are discussing about indoor scenes, object rotations are invariably around the gravity axis. 
Recently, \cite{yu2022rotationally} propose a rotationally equivariant 3D object detector, which is able to detect bounding box that are equivariant to the object pose. But even this work accounts for object rotations along the gravity axis alone. Theoretically, it may be easy to extend the framework in \cite{yu2022rotationally} to SO(3) rotations, but many fundamental issues may need to be solved in practice. Developing a 3D object detector that is equivariant to SO(3) rotations invariably begs the question: ``Are there robust shape descriptors that are rotation-equivariant?". \agp{A recent work \cite{patil2023evaluating} systematically evaluates recent shape classification networks for robustness to rotation invariance, which can provide helpful pointers in designing rotation-in/equivariant architectures for 3D object detection.}

\subsection{3D semantic scene segmentation}
\label{subsec:3d_ssg}
\begin{table*}[!t]
\centering
\resizebox{\textwidth}{!}{%
\begin{tabular*}{1.27\linewidth}{l|c|c|c|c|c|c|c}
\toprule
\textbf{Related Work} & \textbf{\makecell{Learning\\ Framework}} & \textbf{Scene rep} & \textbf{Backbone} & \textbf{Input} & \textbf{Output} & \textbf{Dataset} & \textbf{Evaluation Metric(s)}\\[0.5 ex]
%
\midrule
\cite{qi2017pointnet} & Supervised & RGB-D image & MLP & Scene point cloud & Per-point scores & S3DIS & Acc, mIoU \\
%
\cite{qi2017pointnet++} & Supervised & RGB-D image & PointNet & Scene point cloud & Per-point scores & ScanNet & Acc, mIoU  \\
%
\cite{dai20183dmv} & Supervised & RGB-D image & 2D, 3D CNN & \makecell{Multi-view RGB images\\ and voxel grid} & Per-voxel scores & ScanNet & Acc \\
%
\cite{LiBSWDC18} & Supervised & RGB-D image & CNN & Scene point cloud & Per-point scores & ScanNet, S3DIS & Acc, mIoU \\
%
\cite{thomas2019kpconv} & Supervised & RGB-D image & PointNet (MLP) & Scene point cloud & Per-point scores & ScanNet, S3DIS  & Acc, mIoU \\
%
\cite{wang2019dynamic} & Supervised & RGB-D image & Message Passing MLP & Scene point cloud, point graph & Per-point scores & S3DIS & Acc, mIoU \\
%
\cite{li2019deepgcns} & Supervised & RGB-D image & MLP (GCN) & Scene point cloud, point graph & Per-point scores & S3DIS & Acc, mIoU \\
\cite{ZhaoJFJ19} & Supervised & RGB-D image & MLP (GCN) & Scene point cloud & Per-point scores & ScanNet, S3DIS  & Acc, mIoU \\
%
\cite{ZhouFFW0L21} & Supervised & RGB-D image & MLP (GCN) & Scene point cloud & Per-point scores & S3DIS  & Acc, mIoU \\
%
%
\cite{zhao2021point} & Supervised & RGB-D image & MLP, Transformer & Scene point cloud & Per-point scores & S3DIS & \makecell{Acc, mIoU,\\mean classwise acc} \\
\bottomrule
%
%
\end{tabular*}%
}
\caption{For the task of \textbf{3D indoor semantic scene segmentation} (3D-SSG), we summarize state-of-the-art methods in the table above-- \emph{Input} and \emph{Output} refer to the input consumed by the \emph{Backbone} and its output, respectively. Acc - Accuracy, AP - Average Precision, mAP - mean of Average precision, AR - Average Recall, IoU - Intersection over Union, RMSE - Root of Mean Squared Error.}
\label{tab:3d_scene_seg}
\vspace{-15pt}
\end{table*}


A more in-depth understanding of indoor scenes, beyond 3D object detection, involves segmenting objects based on their semantics. An even comprehensive 3D indoor scene understanding pushes segmentation further, going into \emph{instance} segmentation (not simply semantic). 
\revision{
Indoor scenes contain a variable number of objects that occur in different positions and orientations. Moreover, some subscenes may contain identical sets of similar/identical objects, which can simply be represented by a single model. Determining the spatial extent of objects in an indoor scene (i.e., segmentation) in the presence of different categories with varying number of model instances, geometries, and rotation distributions is quite challenging. This problem also is related to object detection, since for scene segmentation, the underlying models will need to implicitly reason about objects in a scene, allowing for fine-grained localization that result in segmenting the overall object.}  

Most works in the area of 3D indoor scene understanding restrict themselves to semantic segmentation, although there are some works that tackle the instance segmentation problem. Such works primarily build on the intuitions of semantic scene segmentation networks. As such, we cover notable works that propose methods for semantic segmentation, all of which use point clouds as the choice of scene representation. Table \ref{tab:3d_scene_seg} lists notable works on 3D semantic segmentation in indoor scenes.\\
\newline
\begin{figure}[!t]
    \centering
    \includegraphics[width = \linewidth, height = 0.6\linewidth]{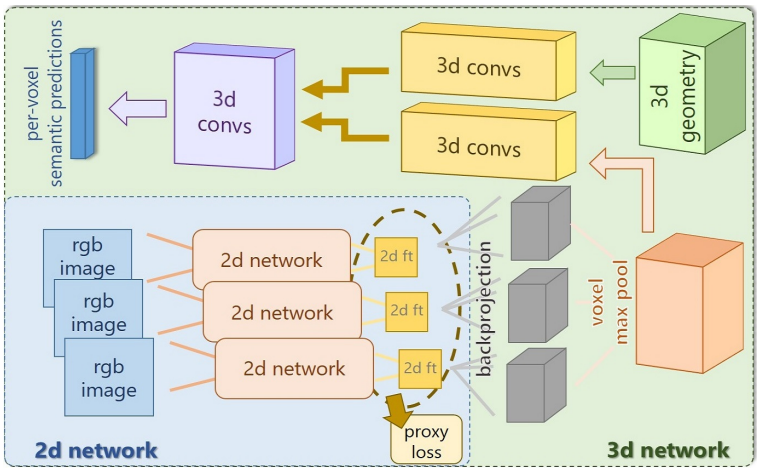}
    \caption{Network architecture for semantic scene segmentation, as proposed in \, cite{dai20183dmv}. It is composed of a 2D network and a 3D one. A 2D CNN extracts features from aligned images of a scene for which a geometric reconstruction is also performed from RGB-D scans. These 2D CNN features are mapped to 3D space using a differentiable back-projection layer. Features from multiple views are max-pooled on a per-voxel basis and fed into a stream of 3D convolutions, along with the reconstructed 3D geometry. Finally, both 3D streams are joined and the 3D per-voxel labels are predicted. The whole network is trained in an end-to-end fashion.}
    \label{fig:3dmv}
    \vspace{-15pt}
\end{figure}

\textbf{Point cloud-based techniques }
Qi et al. \cite{qi2017pointnet} propose the very first point-cloud processing network, PointNet, that can be used for 3D semantic segmentation on shapes and scenes. It has a unified framework (supervised) for object classification and segmentation. By design, the semantic segmentation network of PointNet is an extension of the classification network which takes \emph{n} points as input, applies transformations on the input as well as intermediate feature, and then aggregates point features by max pooling; see Figure \ref{fig:point_net} for an overview. Experiments were performed on the Stanford 3D semantic parsing data which contains 3D scans of indoor environment with semantic annotations per point. During training, random crops with 4096 points coming from a room in the dataset is passed through the network which learns to assign a semantic label to each point (segmentation is essentially a classification task on each point).
During test time, all points forming a room are input to the system and semantic labels are obtained.\\

PointNet++ proposed by Qi et al. \cite{qi2017pointnet++} builds on top of PointNet to further improve semantic scene segmentation. They propose a hierarchical feature learning architecture that uses PointNet as a feature processing block. While PointNet uses a single max pooling operation to aggregate the whole point set, PointNet++ builds a hierarchical grouping of points and progressively abstracts larger and larger local regions along the hierarchy. This leads to better semantic scene segmentation. 
\revision{
However, the max pooling operation used to aggregate features in local neighborhood regions often causes loss of information. PointCNN\cite{LiBSWDC18} proposed an $\chi$-conv operator to adapt the convolutional networks for point clouds. Specifically, to aggregate the local region feature of each point spanned by its K nearest neighbors, the network predicts an $\chi$-transformation matrix to weight and permute the per-point features, which is then processed using element-wise product and sum operations present in a typical convolution operation.
} 
In continuation with point convolutions, Thomas et al. \cite{thomas2019kpconv} introduce \emph{KPConv}, a convolution operation that operates on point clouds, taking radius neighborhoods as inputs and processing them with weights spatially located by a small set of kernel points. A deformable version of this convolution operator is also proposed that learns local shifts, effectively deforming the convolution kernels to make them fit the point cloud geometry. They demonstrated an improved performance on semantic segmentation which could be attributed to the flexibility offered by deformable KPConv.

Wang et al. \cite{wang2019dynamic} propose a supervised dynamic graph convolutional neural network (DGCNN) that uses PointNet as the backbone network, and demonstrate the application of their proposed method for semantic scene segmentation. The key idea is to \emph{compute} point graphs at every layer and applying \emph{edge convolutions} that are invariant to neighbor ordering. Point graphs are computed using k-nearest neighbor (based on L$_{2}$ distance) between points. The dynamic nature of graph computation at every layer of the graph convolution network enables them to capture better local and global features, leading to improved semantic scene segmentation results. Li et al. \cite{li2019deepgcns} propose a dilated version of graph convolution neural networks, which enables them to capture global features better, which is demonstrated by the results of semantic scene segmentation. 

\revision{
In contrast to these sparse graphs connecting the center point and its neighbors, Zhao et al. propose PointWeb\cite{ZhaoJFJ19}, which builds dense, fully connected graphs in local regions and processes them using the Adaptive Feature Adjustment (AFA) module. This module predicts the impact of neighborhood points by adaptively aggregating contextual information on graph edges. Further, Zhou et al. \cite{ZhouFFW0L21} propose an adaptive graph convolution (AdaptConv) that generates adaptive kernels for points within a local region, rather than weighting the features based on fixed/isotropic kernels. On the other hand, Cheng et al. \cite{ChengHXY21} propose SSPC-Net, a semi-supervised method for 3D point cloud segmentation, which partitions the point clouds into super-point graphs, then dynamically propagates information from the super-point labels and uses the coupled attention mechanism to enhance the super-point features for more accurate segmentation.
}
\\

In 3D-SIS \cite{hou20193d}, instance segmentation on 3D scans is performed by learning from both color and geometry input obtained from real RGB-D scans. Specifically, the proposed learning framework has two branches -- one that uses color images corresponding to reconstructed scan geometry, and the second one uses 3D point cloud reconstruction, either chunks of an indoor scene or a full one, from many different frames of RGB-D scan. The backbone for the first branch is a 2D CNN that extracts meaningful color features, which are brought to 3D using a differentiable back projection layer. The second branch uses a 3D CNN to obtain geometry features. Both the color and geometry features are joined in 3D. \\
In order to obtain object masks, they need to be localized. To this end, the work proposes a region proposal network to predict object bounding boxes. From these box predictions, class labels are predicted using a classification head, which are both used for informing instance mask predictions. Note that this is a strongly supervised approach. During inference, instances can be inferred on a full test scene in a single forward pass.

\begin{figure}[!t]
    \centering
    \includegraphics[width = 0.8\linewidth, height = 0.4\linewidth]{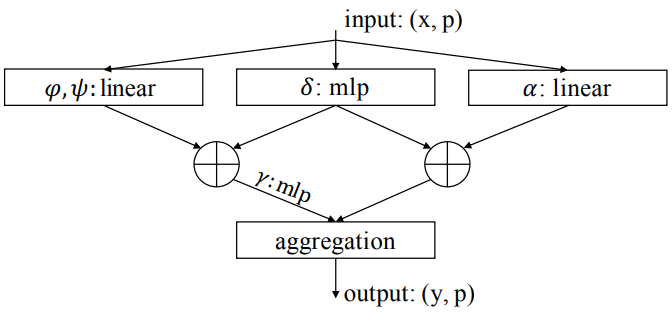}
    \caption{\emph{Point transformer layer} from \cite{zhao2021point}. This module is essentially responsible for extracting meaningful features, which are obtained from a combination of three neural sub-modules as depicted above.}
    \label{fig:point_transformer_layer}
    \vspace{-15pt}
\end{figure}

\begin{figure}[!t]
    \centering
    \includegraphics[width = \linewidth, height = 0.6\linewidth]{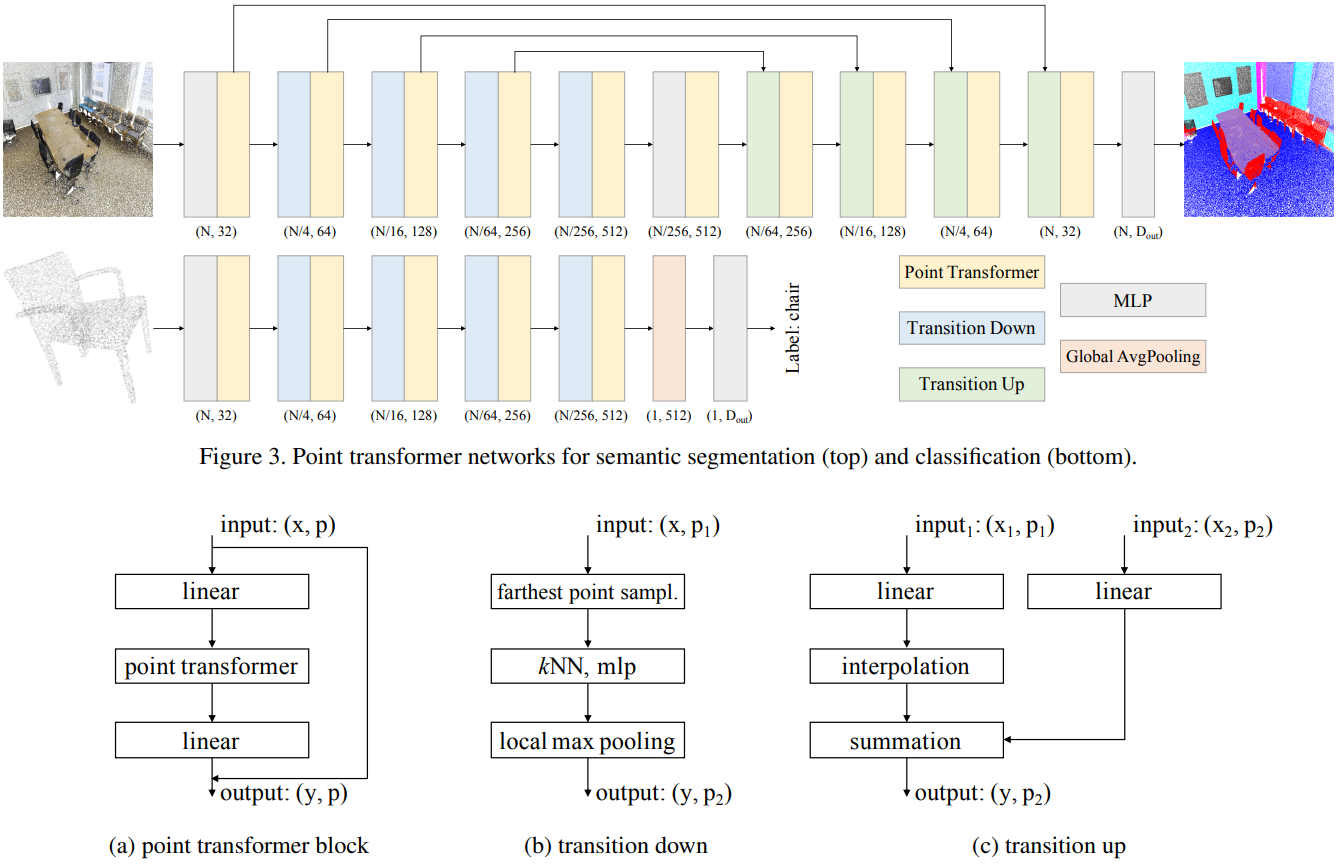}
    \caption{Using the point transformer layer from \ref{fig:point_transformer_layer} as the backbone, \cite{zhao2021point} propose an architecture for point cloud processing tasks such as shape classification, part segmentation and semantic scene segmentation. }
    \label{fig:point_transformer_overall}
\end{figure}

More recently, Zhao et al. \cite{zhao2021point} present a transformer-based architecture that serves as the backbone for many recognition and segmentation tasks on 3D point clouds, including semantic scene segmentation. The key insight driving this work is that the self-attention operator at the core of a transformer is essentially a set operator, and point clouds are essentially sets embedded in metric space. Much like \cite{vaswani2017attention}, their method, called Point Transformer, makes use of encoder and decoder branches which are stacked layers of what they call, a point transformer layer, which is roughly, a piece-wise summation and aggregation of outputs from two different linear layers and an MLP, all of which take points from the point cloud as input; see figure \ref{fig:point_transformer_layer} and \ref{fig:point_transformer_overall}. Although a supervised learning framework like all other works discussed till now, the gain comes from self-attention mechanism that learns the correlation between points in the input point cloud, outperforming state-of-the-art designs including graph-based models, sparse convolutional networks, and continuous convolutional networks.

\textbf{Voxel-based technqiues }
Dai et al. \cite{dai20183dmv} propose a supervised multi-view prediction approach for semantic scene segmentation. The goal of their method is to infer semantic class labels on per-voxel level of the grid of a 3D reconstruction. To achieve this, they propose a 2D-3D neural network that leverages both RGB and geometric information obtained from 3D scans. Their method takes as input a reconstruction of an RGB-D scan along with its color images, and predicts a 3D semantic segmentation in the form of per-voxel labels; see Figure \ref{fig:3dmv} for an overview. To allow for 2D features to influence the per-voxel semantic predictions, they combine the 2D features (2D convolution on multi-view RGB images centered around a voxel location in \emph{xy} plane) with the 3D ones (3D convolutions on volumetric chunks of a scene centered around a point in the \emph{xy} plane) using a differentiable back-projection layer. This joint 2D-3D network is trained in an end-to-end fashion to predict per-voxel classes, resulting in a semantically segmented scene.
\newline

\textbf{Discussion }
With the introduction of PointNet \cite{qi2017pointnet}, there has been an explosion of related works that attempt to solve standard shape+scene understanding tasks -- object classification and part/semantic scene segmentation. It is observed that architectures that are built for the task of semantic scene segmentation are all based on features meant to be used for classification purposes. 
\agp{
For example, point cloud-based architectures for semantic scene segmentation mainly propose different ways to aggregate neighborhood information of each point in a scene, while transformer-based architectures investigate different strategies for improving the attention mechanisms that capture neighborhood relations. 
\\
\newline
Among the different approaches discussed above, point cloud-based methods, which operate directly on the point coordinates, do not suffer from any loss of information that occurs with voxel-based methods that discretize the underlying data into regular grids. However, voxel-based methods capture local neighborhood geometry better than the former, leading to improved segmentation accuracy. That is, compared to point cloud-based methods, they are more context-aware, owing primarily to the spatial/topology awareness of the underlying representation, but are computationally intensive compared to the former. There is thus 
an accuracy vs. efficiency trade-off, which means that the choice of a 3D scene segmentation method to build upon boils down to the complexity of the scene, and the available computational resources. In this context, it is worthwhile to investigate hybrid representations such as \cite{liu2019point, zhang2020deep, zhang2020h3dnet}, which can provide the best of different representations.

However, although the above methods adopt different data representations and propose architectures for 3D semantic scene segmentation task, they all adopt supervised learning frameworks on commonly available datasets, e.g. 3DSIS dataset \cite{hou20193d}. As a result, they tend to perform relatively worse on the categories with low frequency in the training set, with considerable room for improvement in the case of connected and/or overlapping objects in the scenes. Developing semi-supervised or even unsupervised learning techniques for indoor 3D semantic scene segmentation is \emph{the} way to overcome these limitations introduced due to the dataset-specific learning process.

As well, for more general tasks such as robot manipulations (for example, if a robot is tasked to open a cabinet drawer, it has to first understand which parts of the cabinet are interactable, and then localize-plus-segment that specific part on the cabinet), semantic scene segmentation algorithm should not be merely restricted to the knowledge base of training object categories, but an open-vocabulary segmentation scheme should be developed for such general tasks.
}
\\
\newline
All in all, backbone architectures for semantic scene segmentation seem to be pretty matured, with recent efforts being invested in engineering these networks for slightly better performance. A more exciting research direction, one that is directly an extension of semantic segmentation, is \emph{instance segmentation} in 3D scenes. This area allows for fine-grained understanding of scenes since we often encounter sets of identical objects in a scene. For example, a set of chairs surrounding a dining table or a set of place settings on the table. With a wide disparity in the number of instances observed, the distribution over observed rotations, and the geometric variations among instances per model within a category, the challenges are galore. Developing advanced techniques in this direction provides a deeper insight on scene understanding tasks. \\
\newline

\agp{
}

\subsection{3D scene reconstruction}
\label{subsec:scene_recon}
\begin{table*}[!t]
\centering
\resizebox{\textwidth}{!}{%
\begin{tabular*}{1.38\linewidth}{l|c|c|c|c|c|c|c}
\toprule
\textbf{Related Work} & \textbf{\makecell{Learning \\framework}} & \textbf{Scene rep} & \textbf{Backbone} & \textbf{Input} & \textbf{Output} & \textbf{Dataset} & \textbf{Evaluation Metrics}\\[0.5 ex]
%
\midrule
\cite{izadinia2017im2cad} & Unsupervised & RGB image & 2D CNN & \makecell{2D image and \\3D CAD models} & \makecell{CAD models with placements,\\ 3D Room layout} & \makecell{ShapeNet, LSUN,\\ SUN RGBD} & \makecell{Classification Acc,\\ mAP, voxel IoU} \\
%
\midrule
\cite{huang2018cooperative} & Supervised & RGB image & 2D CNN & 2D image & \makecell{3D OBB for object,\\ room layout} & SUN RGB-D & mAP and 2D IoU \\
%
\midrule
%
\cite{nie2020total3dunderstanding} & Supervised & RGB image & 2D CNN with attention & 2D image & \makecell{Room layout OBB, \\3D objects with OBBs} & SUN RGBD, Pix3D & 3D IoU \\
%
\midrule
\cite{zhang2021holistic} & Supervised & RGB image & \makecell{2D CNN with attention,\\ GCN} & 2D image & \makecell{Room layout OBB, \\3D objects with OBBs} & SUN RGBD, Pix3D & 3D IoU \\
%
\midrule
\cite{murez2020atlas} & Supervised & RGB-D image & 2D and 3D CNN & Video/2D image frames & 3D mesh & ScanNet & AP, AR, F-score, RMSE\\
%
\midrule
\cite{avetisyan2020scenecad} & Supervised & RGB-D image & 2D CNN, GNN & \makecell{Scene point cloud,\\ CAD models} & \makecell{Room layout box,\\ CAD models with placements} & SUNCG, ScanNet & 3D IoU, F1 score \\
%
\midrule
\cite{Gkioxari_2022_CVPR} & Unsupervised & RGB image & 2D CNN & RGB image & \makecell{3D objects \\and spatial placement} & \makecell{Scene-ShapeNet, \\HyperSim, ScanNet}  & 2D Box and Mask IoU\\
\bottomrule
%
%
\end{tabular*}%
}
\caption{Notable works on \textbf{3D indoor scene reconstruction} -- \emph{Input} and \emph{Output} refer to the input consumed by the \emph{Backbone} and its output, respectively. We focus on object layout reconstruction over room layouts alone and not on room layout reconstruction. Abbreviations used for evaluation metrics: Acc - Accuracy, AP - Average Precision, mAP - mean of Average precision, AR - Average Recall, IoU - Intersection over Union, RMSE - Root of Mean Squared Error.}
\label{tab:3d_scene_recon}
\vspace{-15pt}
\end{table*}

\revision{
Single-view reconstruction is a severely ill-posed problem, mainly due to the lack of sufficient priors for obtaining a faithful reconstruction. Inferring a 3D structure, either for an object or a scene, from an input image is a complex process that combines low-level image cues to learn the structural arrangement of parts/objects and the high-level semantic object/scene information. Variations in the views and shading, along with variations in textures make conventional reconstruction algorithms fail. For indoor 3D scenes, the focus is more on reconstructing object arrangements than the objects themselves. The challenge here is to reason about 3D positions from a single image, while leveraging contextual information about object arrangements reflected in the input image.
}

In general, indoor 3D scene reconstruction can be categorized into two parts: room layout reconstruction and object layout reconstruction.  Room layout reconstruction deals with recovering the spatial layout of the walls of a room, whereas object layout reconstruction involves recovering the spatial arrangement of 3D objects. We limit the scope to reconstructing object arrangements in this section (there is rich literature on reconstructing the room layouts alone). The input representations for 3D scene reconstruction tasks span the entire spectrum of visual representations, the most prominent ones being RGB (D) images and point clouds. We cover 3D object layout reconstruction works from a single RGB image \cite{izadinia2017im2cad, nie2020total3dunderstanding, zhang2021holistic, huang2018cooperative, huang2018holistic, denninger20203d}, or a set of posed RGB images \cite{murez2020atlas}, or a point cloud scan of an indoor scene \cite{song2017semantic}, as summarized below and in Table \ref{tab:3d_scene_recon}.

\begin{figure}[!t]
    \centering
    \includegraphics[width = \linewidth, height = 0.5\linewidth]{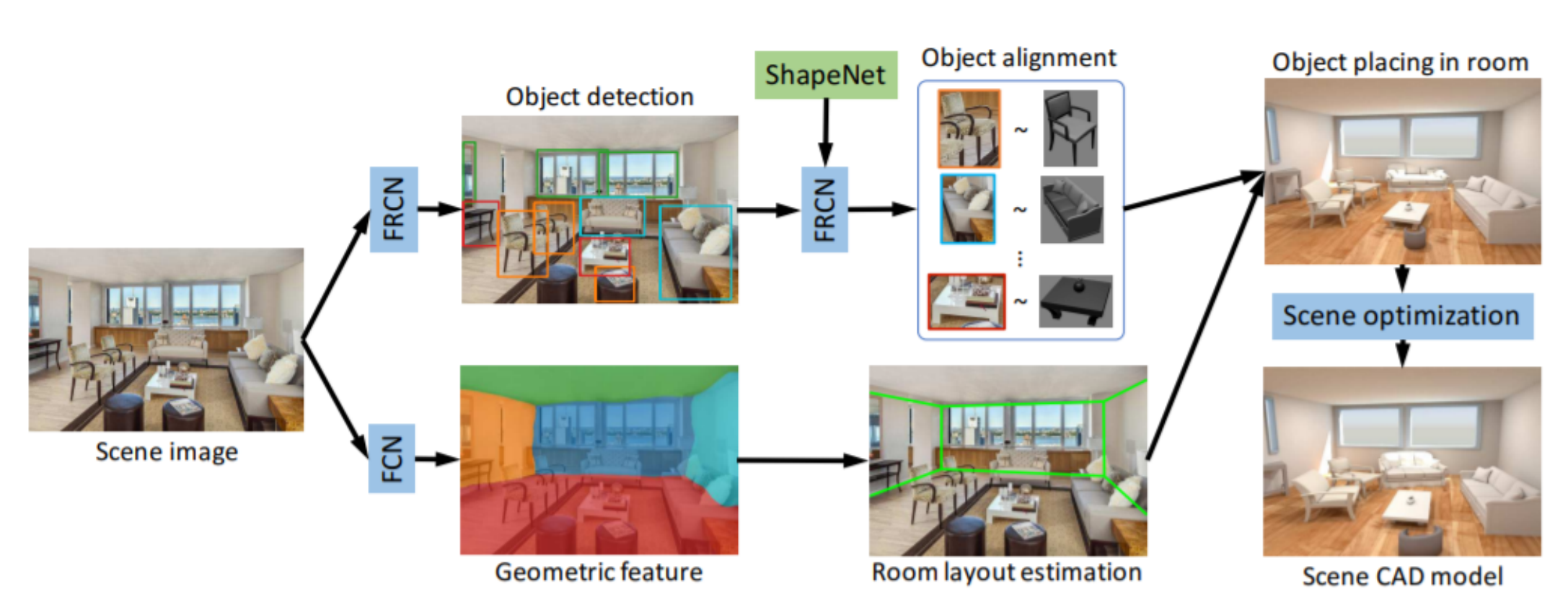}
    \caption{\emph{IM2CAD} \cite{izadinia2017im2cad} proposes a system to reconstruct a CAD modeled scene based on a single input image. The main idea is to first render CAD models from different viewpoints and match their CNN features to that of detected objects in the input image. Once CAD models are retrieved corresponding to objects in the input image, an optimization algorithm modifies places them in a scene to reflect object arrangement in the input image. Color and lighting are injected using an additional module.}
    \label{fig:im2cad}
    \vspace{-15pt}
\end{figure}

\textbf{Image-based techniques }
Izadania et al. \cite{izadinia2017im2cad} propose a method for reconstructing a 3D scene of an RGB image (see Figure \ref{fig:im2cad}). They make use of a pre-trained 2D object detector (FasterRCNN \cite{ren2015faster}) to detect objects in the input RGB image, and compare the box features of these 2D  detections with that obtained from multi-view renderings of a database of CAD models (ShapeNet). This comparison enables them to retrieve an approximately aligned CAD model for a detected object in the input image. This process is done for all the object categories in consideration (eight, to be precise). 
In parallel, a fully convolutional network (FCN) \cite{long2015fully} is trained for room layout estimation that estimates per-pixel surface labels for ceiling, floor, and walls. The FCN network is trained on annotated scenes from the LSUN database \cite{yu2015lsun}.

In order to find the object location and scale in the x and y directions (i.e., parallel to the ground plane) in the scene, a ray is cast from a camera center through the input image pixels corresponding to the bottom four corners of an aligned CAD model cube (note that all CAD models in ShapeNet dataset are confined within a unit cube). 
To compute the object scale along the z axis, they compute the ratio between the length of the four vertical edges of the projected cube and the length of those edges from the ground plane to the intersection of those lines with the horizontal vanishing line. 
After estimating the 3D room geometry and the initial placement of the objects in the scene, object placements are refined by optimizing the visual similarity of the
rendered scene with that of the input image. To this end, they solve an optimization problem where the variables are the 3D object configurations in the scene and the objective function is the minimization of the cosine distance between the convolutional features obtained from the camera view rendered scene and the input image.\\

\begin{figure}[!t]
    \centering
    \includegraphics[width = \linewidth, height = 0.5\linewidth]{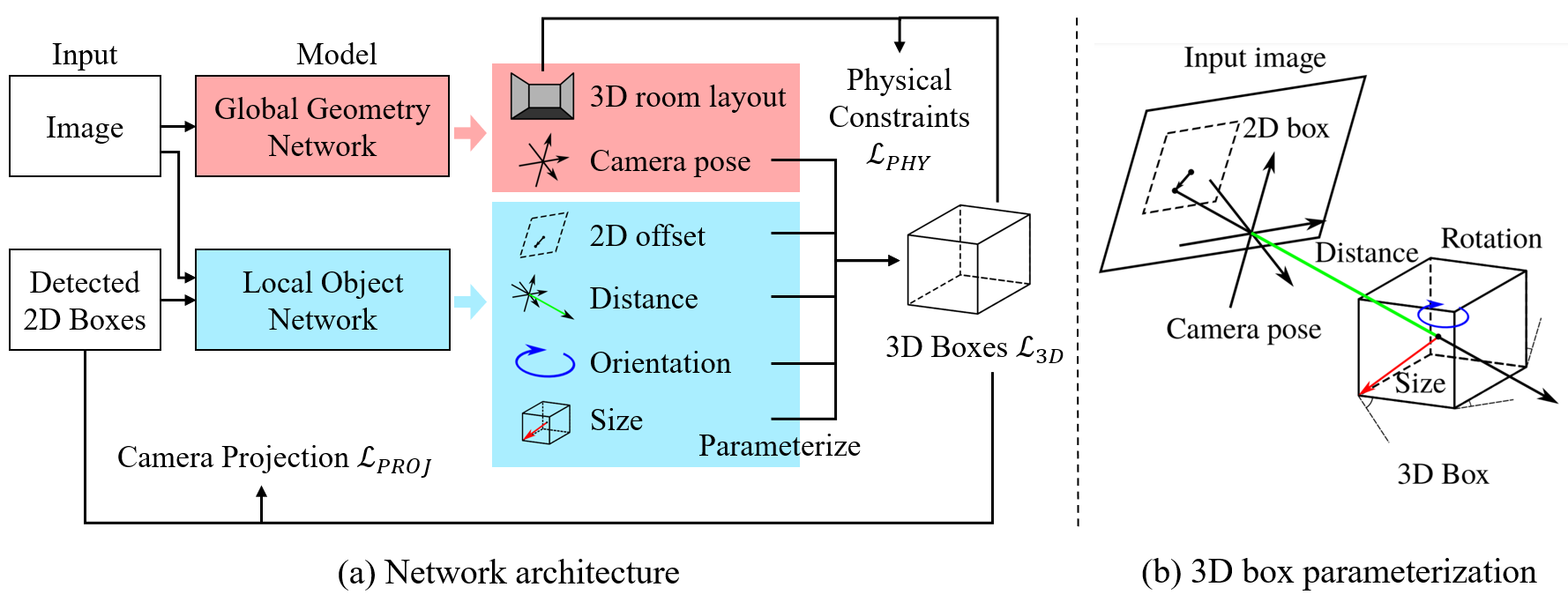}
    \caption{Single-view scene reconstruction pipeline from \cite{huang2018cooperative}. The method proposes to cooperatively estimate 3D object bounding boxes, layout bounding box and the camera pose, and project the resulting 3D layout to image plane forcing consistency between the input image and the
    projected image.}
    \label{fig:coop_sun}
    \vspace{-15pt}
\end{figure}

Huang et al. \cite{huang2018cooperative} use a strongly supervised approach for reconstructing a 3D scene for a given RGB image, in what they call, a "cooperative" manner. Instead of developing independent modules for reconstructing parts of the scene, they propose to cooperatively estimate 3D object bounding boxes, layout bounding box and the camera pose, and project the resulting 3D layout to the image plane forcing consistency between the input image and the projected image, see Figure \ref{fig:coop_sun}. Specifically, the following three cooperative losses are used: 3D box loss (\emph{directly} optimize final estimation of the 3D boxes), 2D projection loss (maintain the \emph{consistency} between 2D image and estimated 3D boxes) and physical loss (penalize the \emph{physical} violations between 3D objects and 3D room layout). This kind of cooperation is shown to improve the estimation accuracy of 3D bounding boxes, and the physical plausibility of the overall scene. Supervisory signals are obtained from the SUNRGB-D dataset \cite{song2015sun}.

\begin{figure*}[!t]
    \centering
    \includegraphics[width = \linewidth]{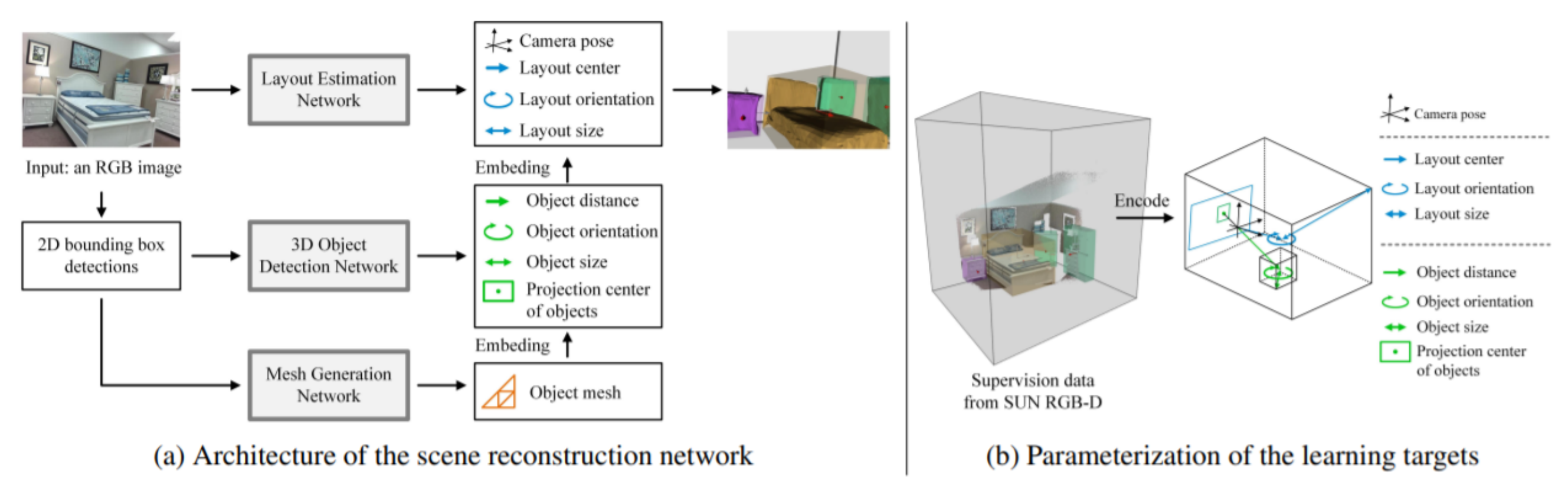}
    \caption{Figure illustrates the pipeline for reconstructing a 3D scene from a single image in a supervised setting. The work is called \emph{Total3DUnderstanding} \cite{nie2020total3dunderstanding}, or \emph{T3DU} in short. Relationships between detected objects in the input image are captured using attention mechanism. Both object and room layout are recovered and represented in the form of a cuboid box. Objects are reconstrcuted using a mesh generation network from \cite{groueix2018papier}.}
    \label{fig:t3du}
    \vspace{-15pt}
\end{figure*}

Similar to \cite{huang2018cooperative}, Nie et al. \cite{nie2020total3dunderstanding} propose a supervised method to jointly reconstruct room layout, object poses and meshes in an indoor scene from a single RGB image as the input, termed as \emph{Total3DUnderstanding} (T3DU). Their approach consists of three parts: room layout estimation (in world coordinate system), 3D object detection (in camera coordinate system), and mesh generation (in object canonical system). The output of these three modules are embedded together in the reverse order (see Figure \ref{fig:t3du}), to establish an end-to-end joint training mechanism. The 3D object detection module makes use of an attention mechanism to obtain contextual object features (called "Relational features") for a detected 2D object in the input image. The relational features are combined with the detected 2D object features (obtained using a ResNet), and the resulting features are regressed through an MLP to get 3D bounding boxes. Room layout estimation is done similarly, where the room bounding box parameters are regressed using the 3D object detector module. Finally, a mesh generation module based on AtlasNet \cite{groueix2018papier} is employed to reconstruct 3D object meshes.  

During inference, the generated meshes in the canonical system are transformed to the camera system for viewing object bounding boxes, which are in turn converted into world coordinate system for combined interpretation with the estimated room layout bounding box. This method of training in an end-to-end fashion produced improvements over \cite{huang2018cooperative} on room layout estimation and 3D object detection, and this gain can be attributed to the incorporation of relational features in 3D object detection and room layout estimation, that take scene context into account via the attention mechanism.
\\

\begin{figure*}[!t]
    \centering
    \includegraphics[width = 0.8\linewidth, height = 0.25\linewidth]{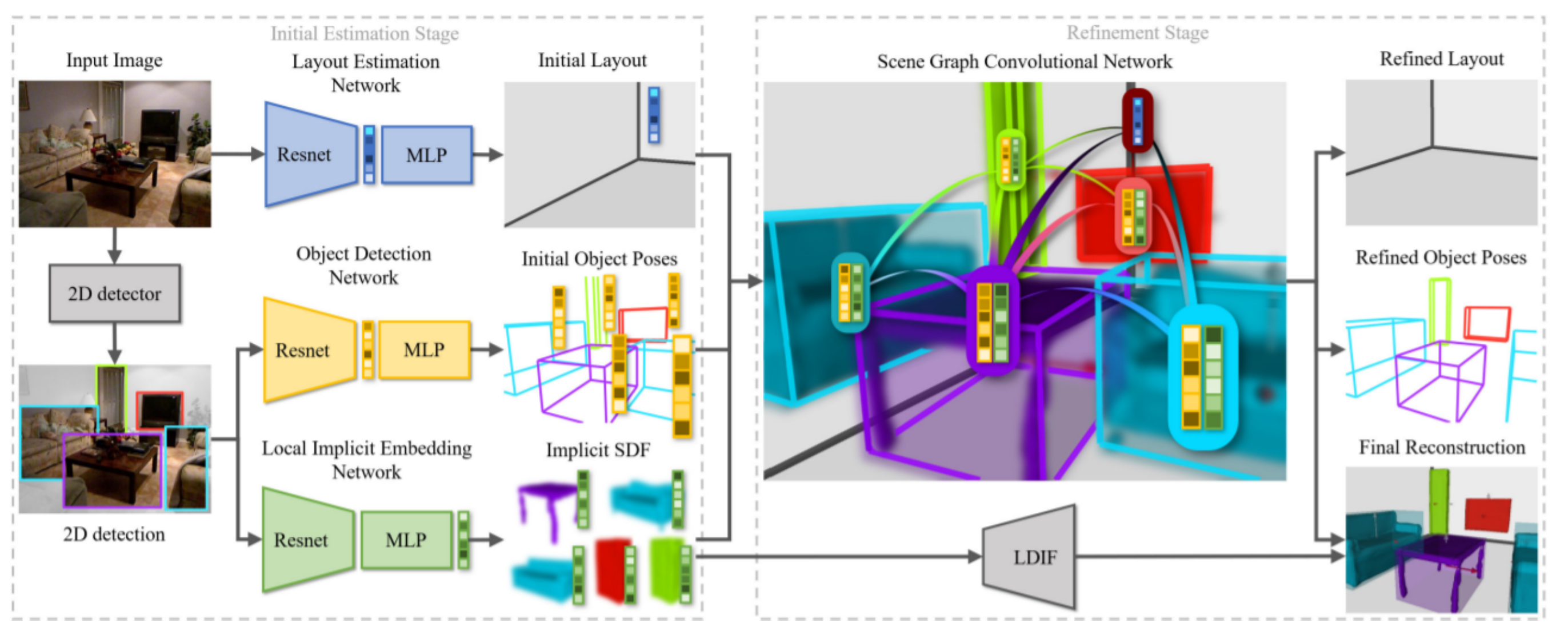}
    \caption{Figure illustrates various modules present in reconstructing a 3D scene from a single image as proposed in \cite{zhang2021holistic}. It is a supervised learning framework. The work mainly builds on Total3DU \cite{nie2020total3dunderstanding}, with the major difference being the use of implicit representations for objects in a scene. In addition, a graph neural network is employed on the scene objects, allowing to capture better contextual information. Room and object layout are represented using box cuboids, and objects are reconstructed using LDIF \cite{genova2020local}.}
    \label{fig:hol_t3du}
    \vspace{-15pt}
\end{figure*}

Building on T3DU, Zhang et al. \cite{zhang2021holistic} make improvements in the quality of generated object meshes, and reduce cases of object intersections observed in T3DU. In terms of network architecture, the method is split into two stages: initial estimation stage, and a refinement stage; see Figure \ref{fig:hol_t3du}. The initial estimation stage is the same as T3DU (Figure \ref{fig:t3du}(a)), with the only difference being the employment of a network outputting a shape code based on local implicit fields \cite{genova2020local}, instead of using AtlasNet \cite{groueix2018papier} for mesh generation. In the refinement stage, the scene is modeled as a graph to capture the scene context, where the node features are concatenation of features obtained from the three modules in the initial estimation step. The node features are updated using message passing, which are later decoded into residuals to refine the initial estimation. The refined poses are then incorporated with the object shapes decoded from shape code with LDIF \cite{genova2020local} to get the final reconstruction of the whole scene. The results show an improvement over existing relevant works which can be attributed to: (a) an improved mesh generation network, (b) the incorporation of a loss term that penalizes the physical interaction of objects (in addition to other losses) and (c) modeling scenes as graphs that helps to better capture scene context and in turn, can effectively refine the initial estimates.

\begin{figure}[!t]
    \centering
    \includegraphics[width = \linewidth]{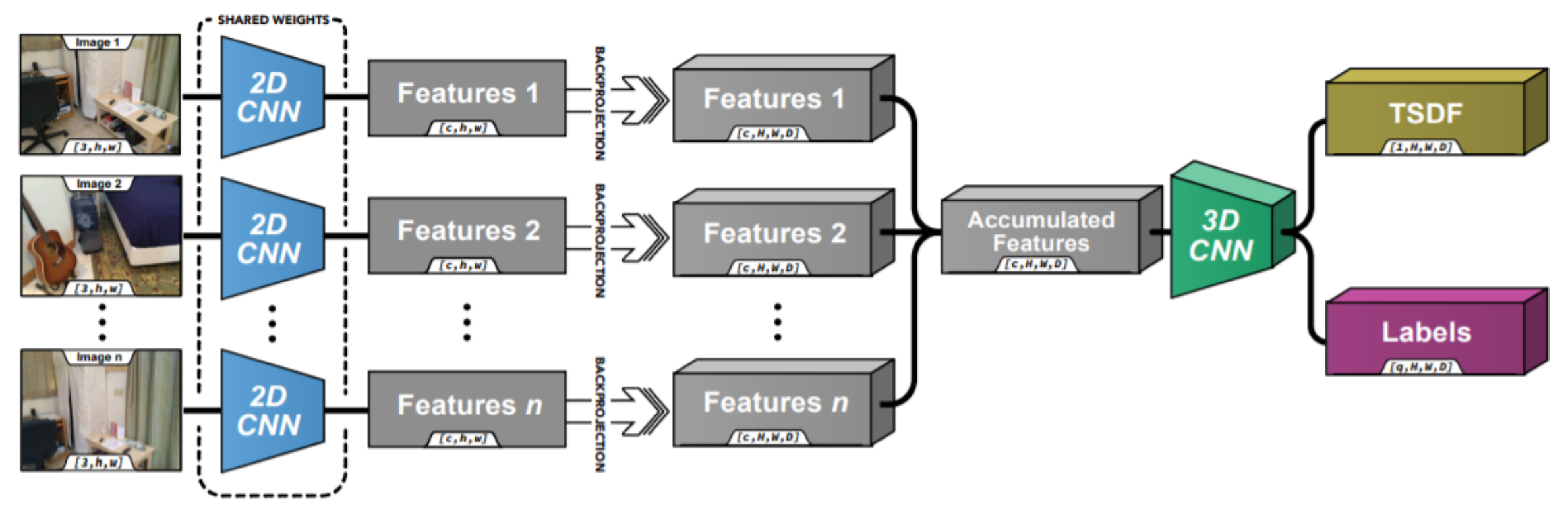}
    \caption{End-to-end training pipeline for 3D scene reconstruction from posed images \cite{murez2020atlas}. The method takes a monocular video as input and extracts 2D CNN features from each frame, which are back-projected to 3D voxel grids using known camera information. A 3D CNN refines the features and regresses a TSDF function, from which a scene mesh is extracted using the marching cubes algorithm.}
    \label{fig:atlas}
    \vspace{-15pt}
\end{figure}

Different from the aforementioned approaches, Murez et al. \cite{murez2020atlas} present \emph{Atlas}, an end-to-end 3D scene reconstruction approach from \emph{posed images}. The method takes a calibrated monocular video as input. Image features from each frame are extracted using a 2D convolutional neural network. These features are back-projected along rays into a 3D voxel volume using known camera intrinsics and extrinsics. Feature volumes are accumulated using a simple running average. After accumulation, a 3D convolutional neural network refines the features and regresses a truncated signed distance function (TSDF). The overall approach is shown in Figure \ref{fig:atlas}. Finally, a mesh is extracted from the TSDF volume using marching cubes. Additionally, semantic labels can be predicted by adding a classification head to the 3D CNN. This approach demonstrated superior quality of reconstructions on challenging long-frame temporal sequences with unobserved geometry, despite not making use of any depth information.

A more recent work called USL \cite{Gkioxari_2022_CVPR} presents an approach to scene reconstruction without any layout supervision, albeit from multi-view images of a scene during training. Their proposed system models object shapes and scene layout to roughly mimic an input image during test time. During training, given two views of a scene, one being the input to the system and the other being the target, a 3D scene is produced for the input view based on the prior work of MeshRCNN \cite{gkioxari2019mesh}. This 3D scene is now rendered from the viewpoint of the target image, and an optimization algorithm compares object masks in the two. This helps in optimizing object arrangements thereby improving the scene layout. 

\textbf{Scan-based methods }
In SceneCAD \cite{avetisyan2020scenecad}, a solution to scene reconstruction from RGB-D scans is proposed. The problem is broken down to that of aligning CAD models to objects in RGB-D scans. The prelude to SceneCAD is a prior work called Scan2CAD \cite{avetisyan2019scan2cad} that aims at estimating object arrangements from 3D scans by learning to align CAD models to RGB-D scans. The emphasis here is on the type of input -- RGB-D scans, which tend to be very noisy and incomplete, with no semantic information. SceneCAD solves a joint problem of estimating both object and room layout information, by capturing relationships among objects and between objects and room elements, such as walls. The end result is a lightweight digitized representation for the input RGB-D scan. Note that this is also a supervised learning approach, with supervision on class labels during object mask predictions and edge labels between graph nodes. 
Many recent approaches to single-view scene reconstruction \cite{kuo2020mask2cad, kuo2021patch2cad, gumeli2022roca} propose similar proxy solutions based on shape recovery of detected 2D objects in the input image via CAD model alignment.

\textbf{Discussion }
3D supervision is hard to obtain, which necessitates the exploration of challenging tasks to scene reconstruction from one or more images. In order to truly reconstruct an indoor 3D scene, both room layout (walls and their arrangement), and object layout (spatial placements of objects in a room) should be recovered, which are non-trivial to solve. In 3D vision, a lot of effort has been put in recovering room layouts from a single image, starting from \cite{hedau2009recovering} to more recent works like \cite{dasgupta2016delay, lee2017roomnet, zou2018layoutnet, huang2018holistic, yang2019dula}. Model-based room layout reconstruction algorithms such as the ones proposed by \cite{hedau2009recovering, mallya2015learning} are limited by hand-crafted features based on one/few properties of physical existence. Learning from data, however, can uncover the full potential of realizable results, allowing to explore further in that direction.

For object layout reconstruction from single/multiple images or 3D scans, two major approaches exist -- (1) learning contextual placements of objects, or (2) aligning CAD models to the input. The first approach is mainly a supervised setting, based on attention mechanism \cite{nie2020total3dunderstanding} or message passing \cite{zhang2021holistic} since they need a notion of what plausible placements look like. In addition to recovering object layouts, these works also perform reconstruction at the object level, which may not be ideal for visualizing results since single-view reconstruction for objects is a research area in itself. A more recent work \cite{Gkioxari_2022_CVPR} does not make use of layout supervision, but gains additional training information by making use of \emph{multi-view} images of the same scene, which compensates for the lack of layout-level supervision. 

The second approach is an ad-hoc setting, moving towards unsupervision at the layout level, where the goal is to simply align CAD models from a database to detected objects in the input image \cite{izadinia2017im2cad} or input 3D scans \cite{avetisyan2020scenecad, avetisyan2019end, avetisyan2019scan2cad, kuo2020mask2cad, kuo2021patch2cad, gumeli2022roca, maninis2022vid2cad}. These approaches are more closer to works that estimate object poses from a single input image, with an added component of optimizing for object placements that reflect the input image/scan. Overall, an unsupervised learning framework that could be trained from a single input image is missing.



\subsection{3D scene similarity}
\label{subsec:scene_sim}

In visual computing, a metric is used to compare different data representations such as two images, meshes, voxels etc., and provide a measure of closeness or similarity between samples in consideration. As a result, they find applications in database retrieval, data clustering, and evaluating the diversity of generative models.

Similarity metrics for 3D shapes (Chamfer Distance, IoU, Light Field Distance etc.) and 2D images (L2 distance, PSNR) make an underlying assumption that the data being compared can be globally aligned.
However, the concept of global alignment between two 3D scenes, even if they are of the same type, is rather weak since there is no "correct" sequence of populating 3D objects in a space to compose a plausible scene, where same objects can be placed quite differently in two scenes. In addition, scene comparison is complicated when the two scenes (of the same type) have different 3D objects, both semantically and geometrically. As such, developing a metric for comparing 3D scenes is quite challenging, but interesting at the same time due to many degrees of design freedom. 
\newline

\begin{figure}[!t]
    \centering
    \includegraphics[width = \linewidth]{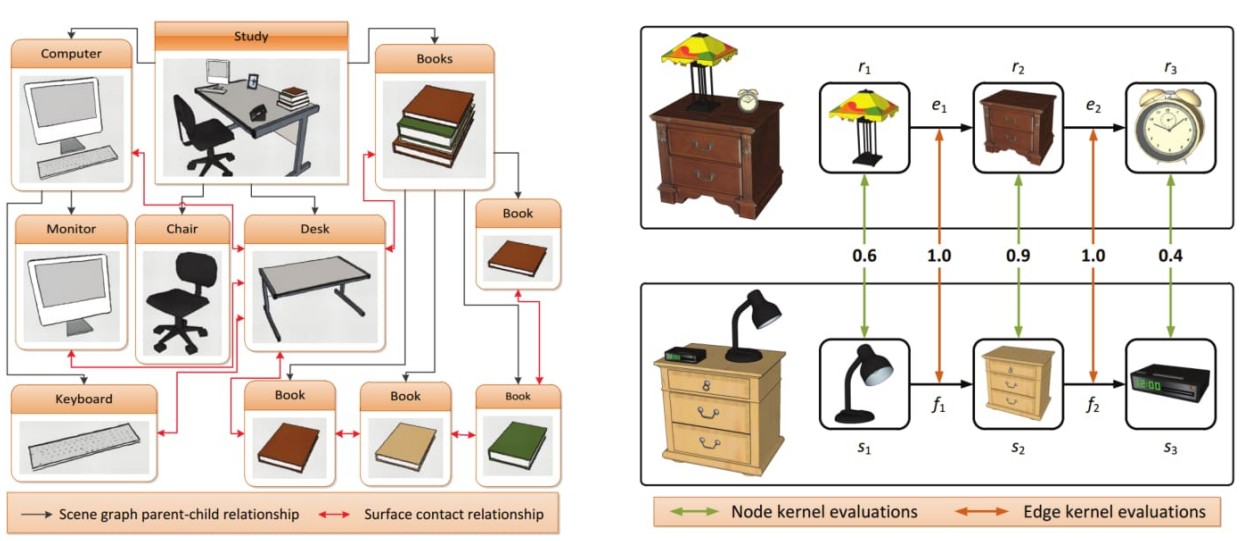}
    \caption{\emph{Graph Kernels} in \cite{fisher2011characterizing} is a method for characterizing the structural relationships between two subscenes. In the figure above, a scene is represented as a relational graph shown on the left. Two types of relationships are considered, as indicated by the arrows. The figure on the right shows an example of the process involved in finding similarity between two subscenes using graph walks. Both walks in each scene are rooted at the lamp
    node. The two walks are compared by taking the product of kernel evaluations for their constituent nodes and edges.}
    \label{fig:GK}
    \vspace{-15pt}
\end{figure}
\textbf{Graph Kernels }
Fisher et al. \cite{fisher2011characterizing} was the first work to develop a method to measure 3D scene similarity, called \emph{Graph Kernels}. Specifically, they characterize 3D scenes using graphs (see Figure \ref{fig:GK} left), where the edges of a graph encode physical proximity relationships (such as support, contact, enclosure) between objects (nodes) in a scene and the nodes correspond to objects in the scene, which encode the geometric information of the objects. With such graph-based representation of scenes, a kernel is defined for comparison of two relationship graphs in a way that similarities between the graph nodes and edges are computed and accumulated to produce an overall similarity of two graphs (see Figure \ref{fig:GK} right). \\
\begin{figure}[!t]
    \centering
    \includegraphics[width = \linewidth]{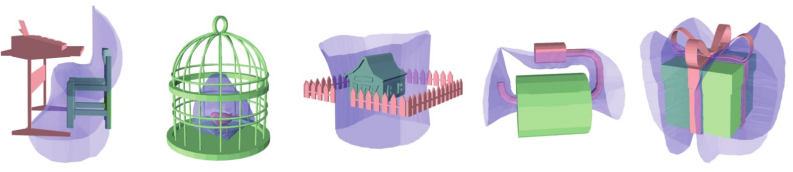}
    \caption{\emph{Interaction Bisector Surface (IBS)} \cite{zhao2014indexing} is a rich representation between objects in a scene that describes topological and geometric relationships between objects in a scene. IBS is the set of points equidistant from two sets of points sampled on different objects, shown as the blue-colored surface above.}
    \label{fig:IBS-a}
    \vspace{-15pt}
\end{figure}

\begin{figure}[!t]
    \centering
    \includegraphics[width = \linewidth]{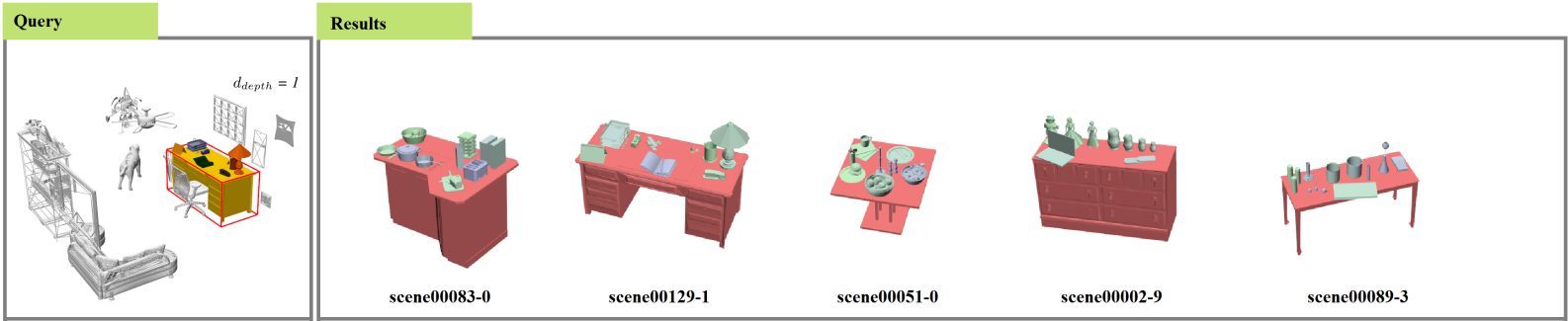}
    \caption{Retrievals results using IBS features \cite{zhao2014indexing} -- In the query scene on the left, a desk, overlayed using its bounding box, is the query object for IBS algorithm. On the right are the scenes ordered based on their similarity, where the red object is the retrieved object with a similar context to the query desk.}
    \label{fig:ibs retrievals}
    \vspace{-15pt}
\end{figure}

\textbf{Interaction descriptors }
Zhao et al. \cite{zhao2014indexing} propose a scene relationship descriptor, called Interaction Bisector Surface (IBS), to characterize complex relationships in a scene, or rather, a sub-scene. IBS describes topological (wrapped in, linked to or tangled with) as well as spatially proximal relationships between objects (see Figure \ref{fig:IBS-a}). IBS is defined as the set of points that are equidistant from two objects, which form an approximation of the Voronoi diagram for objects in the scene.  IBS is used to define a similarity metric between objects, which is then used to group similar objects in a bottom-up manner to automatically construct hierarchies. Unlike \cite{fisher2011characterizing}, IBS does not make use of object labels, and instead, focuses on modeling interaction between objects where spatial relationships between objects are characterized by topological and geometric features. Thus, IBS enables content-based relationship retrieval based on interaction similarity. Figure \ref{fig:ibs retrievals} provides an example of content-based relationship retrieval based on IBS. \\

\begin{figure}[!t]
    \centering
    \includegraphics[width = \linewidth]{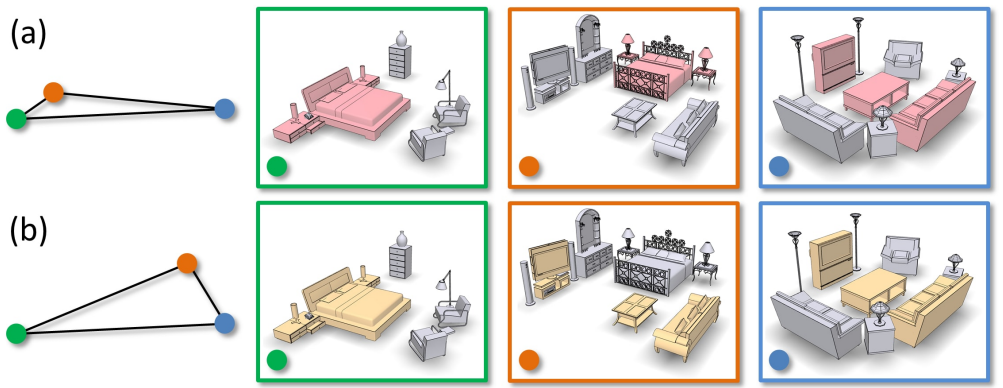}
    \caption{\cite{xu2014organizing} introduce \emph{focal points}, shown in yellow and pink color in scene renderings above, to analyze and organize 3D indoor scenes. Focal points are essentially sub-scenes. The triangles on the left provide a visual illustration for similarity distances between scenes based on focal points.}
    \label{fig:focal points a}
    \vspace{-15pt}
\end{figure}

\begin{figure}[!t]
    \centering
    \includegraphics[width = \linewidth]{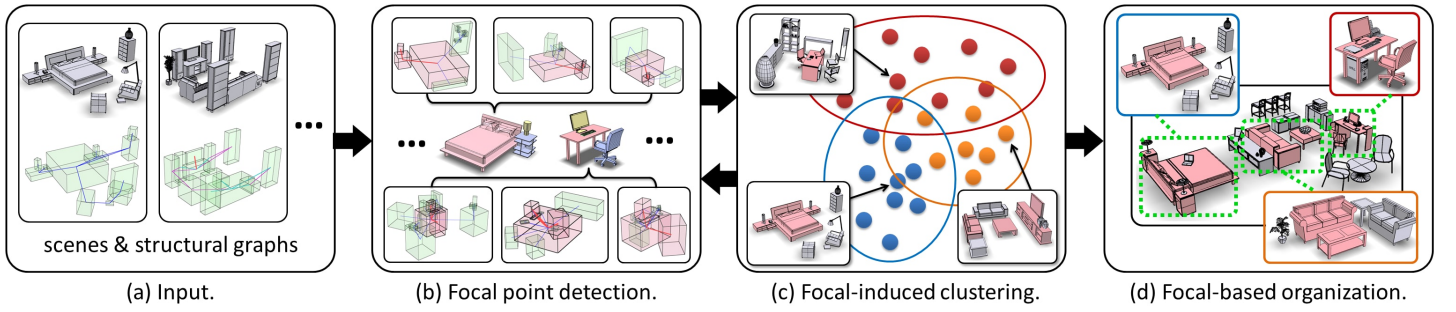}
    \caption{In \emph{focal-cenric graph kernel (FCGK)} \cite{xu2014organizing}, the input to the system is a non-uniform collection of 3D scenes, where each scene is represented by a structural graph (left). The proposed method performs a co-analysis on the collection of 3D scenes to obtain a set of contextual focal points and is an iterative process (middle two). Once focals are obtained, the entire scene collection can be organized with reference to these focals, serving as the interlinks between scenes from various clusters (right).}
    \label{fig:focal points b}
    \vspace{-15pt}
\end{figure}

\textbf{Object-centric descriptor }
Xu et al. \cite{xu2014organizing} present a method to organize a heterogeneous collection of scenes by what they call \emph{focal points}. The key insight in this work is that analyzing complex and heterogeneous scenes in a collection is difficult without references to certain points of attention or focus. \emph{Focal points} provide such points of attention against which two or more complex scenes could be compared. 
Specifically, \emph{focal points} are defined as representative substructures in a scene collection, using which similarity distances between scenes could be computed, see Figure \ref{fig:focal points a} for an illustration. Identifying focal points from a collection of scenes is a problem that is coupled with clustering scenes based on a common point of reference. To solve the coupled problems of focal
point extraction and scene clustering, a co-analysis algorithm that interleaves frequent pattern mining and subspace clustering is presented to extract a set of contextual focal points that guide scene clustering from within the collection. This is shown in Figure  \ref{fig:focal points b}. This co-analysis-based method of focal point extraction extends itself towards scene comparison (retrieval), and exploration of a heterogeneous collection of scenes. 

\textbf{Learning on scene graphs }
Recently, \cite{wald2020learning} propose a neural network that infers a semantic scene graph from an instance-segmentation of a 3D scene, represented as a point cloud. Leveraging these \emph{learned} semantic scene graphs, a 3D scene retrieval is performed where the objects in the scene represent the nodes in the learned graph, and edges represent generic connection as well as semantic relations (such as: next to, lying on, close by) between scene objects. This is achieved by graph matching, not using any neural network, but using a deterministic similarity function based on two types of metric -- Jaccard coefficent and Szymkiewicz-Simpson (SS) coefficient. When matching two graphs G and G$^{'}$, they combine the similarity metric of the object semantics, generic
node edges E as well as semantic relationships R. 

Since the semantic graphs are rich with relational semantics between objects, and the similarity function based on either the Jaccard coefficient or the Szymkiewicz-Simpson coefficient can provide meaningful similarity scores, especially SS coefficient when the two scenes A and B have very different sizes, one can use this retrieval method to find rooms that fulfill certain requirements such as the availability of objects e.g. meeting room with a TV, whiteboard.

\textbf{Discussion}
Efficient retrieval of 3D scenes is helpful in visualizing interior design possibilities, and \agp{a reliable scene similarity metric is central to this application, which can also help with evaluating different scenes}. The query to such evaluation systems can be in different forms -- 2D image, text input, a sketch, a scene graph, a sequence of attributes, etc., each of which poses unique challenges. The central goal to this problem is to define a similarity function that captures the space of scenes based on either pre-defined properties (such as focal-centric themes) or incorporates as many attributes of a scene as possible directly from the data. 

\agp{We observe a clear shift from heuristic-based methods to learning-based methods in formulating a scene similarity measure. This is expected since it is rarely tractable to describe all kinds of scenes using a specific set of rules, especially considering the complexity of scene structures in indoor environments. In the context of learning-based methods,} as discussed above, graph neural networks (GNN) appear to sit comfortably in becoming the preferred tool to tackle 3D scene retrieval problems. In recent years, \emph{neural} graph matching networks \cite{patil2021layoutgmn} have been shown to effectively capture similarity on 2D layouts, albeit with inherent sluggishness due to dependent graph embeddings (embedding of one graph is equally governed by the other graph in a pair). This could be extended to 3D scenes, where the focus should be on efficiently matching scene graphs in a large database, using hashing techniques. \agp{It is worthwhile to explore the assimilation of human feedback in the development of a scene similarity metric since it can help provide a human perspective to an objective algorithm that may not always see things as we humans do.}

\section{3D scene synthesis}
\label{sec:synthesis}

Real world 3D scenes are realized from sequential placement and adjustment of objects carried out in a region-bounded space. Such object placements follow certain interior design rules based on room functionality and layout, which provide useful priors for developing algorithms for modeling indoor 3D scenes. 

Before deep learning made inroads in this field \cite{wang2018deep, li2019grains}, many scene synthesis works \cite{yu2011make, merrell2011interactive, fisher2012example, fisher2015activity, ma2016action, ma2018language, savva2016pigraphs, yu2015clutterpalette} were model-driven and learned from a few hundred 3D scenes. They were progressive in nature (vs. \emph{auto-regressive} terminology used with neural-based works), i.e., the placement of a new object in the scene is conditioned on either one or a set of already existing objects in the scene thus far. These methods are example-based approaches, i.e., the underlying method for scene modeling depends on a set of scene/sub-scene examples to learn priors from. 

We cover notable works on scene synthesis that incorporate different forms of representation discussed in Sec \ref{sec:layout_reps}. We borrow some pointers from the survey on generative models for structured scenes \cite{chaudhuri2020learning} for our report, albeit it is less up-to-date and focuses only on structural representation. \\

\textbf{Probabilistic synthesis}
Relying on probabilistic reasoning over scene exemplars forms the core of example-based scene synthesis approaches \cite{merrell2011interactive, fisher2012example, jiang2012learning, savva2016pigraphs, zhao2016relationship}. Notably, Fisher et al. \cite{fisher2012example} develop a Bayesian network for object co-occurrences and model object placements using a Gaussian mixture model (GMM). To synthesize new scenes from an example scene, object contextual placements are sampled from the learned GMM model. Most of the probabilistic scene synthesis algorithms follow this paradigm, but with variations on both heuristics and models capturing object co-occurrences and their relationships.\\

\textbf{Progressive synthesis}
Models that synthesize a scene by sequential placements of an object or a set of objects (called sub-scene) are said to be progressive in nature. This is a reflection of how scenes evolve in the real world -- based on human actions, which in turn, depend on the functionality of objects present in the scene. This forms the basis of most human-centric scene synthesis works such as \cite{ma2016action, savva2016pigraphs}, discussed later below. In other words, progressive scene synthesis involves user input in some form, be it activity-driven or language-driven \cite{ma2018language}. Such methods can even be made interactive offering more controllability, where the overall system is localized at every synthesis step.

The first instance of such progressive synthesis via interactive modeling was demonstrated by Merrell et al. \cite{merrell2011interactive}. They developed a modeling tool for furniture layout arrangement based on interior design guidelines. The design guidelines are encoded as terms in a probability density function and the suggested layouts are generated by conditional sampling of this function. Another related work called \textit{Make It Home} \cite{yu2011make} offers an interactive modeling tool to synthesize furniture layouts by optimizing a layout function that encodes spatial relationships between furniture objects. \textit{ClutterPalette} \cite{yu2015clutterpalette} presents another interactive modeling tool that progressively populates a scene by suggesting a set of possible objects, the priors of which are learned from the data, when a user clicks on a particular region of the scene.  \newline
%
\begin{table*}[!t]
\centering
\resizebox{\textwidth}{!}{%
\begin{tabular*}{1.38\linewidth}{l|c|c|c|c|c|c|c}
\toprule
\textbf{Related Work} & \textbf{\makecell{Learning \\framework}} & \textbf{Scene rep} & \textbf{Backbone} & \textbf{Input} & \textbf{Output} & \textbf{Dataset} & \textbf{Evaluation Metrics}\\[0.5 ex]
%
\midrule
\cite{li2019grains} & Self-supervised & Hierarchy & RvNN-VAE (MLP) & \makecell{Scene hierarchy \\with object labels,\\ 3D bounding box, \\relative position \\between two siblings} & Scene hierarchy & SUNCG &  \makecell{Perceptual studies, \\Object co-occurrence map}\\
%
\midrule
\cite{wang2018deep} & Self-supervised & 2D image & 2D CNN, MLP & C+6 channel image & $Pr^{(x,y,z,\theta)}$(C) & SUNCG & Perceptual studies \\
%
\midrule
\cite{wang2019planit} & Self-supervised & 2D image, Graph & 2D CNN, GNN & \makecell{Scene graph,\\ C+6 channel image} & $Pr^{(x,y,z,\theta)}$(C) & SUNCG &  \makecell{Perceptual studies,\\ Real/synthetic classifi\\-cation accuracy} \\
%
\midrule
\cite{zhang2020deep}  & Supervised & \makecell{Top-view scene image,\\ object matrix} & \makecell{2D CNN,\\ linear layers} & \makecell{RGB image, \\$n\times$($k$+9) scene matrix} & $n\times$($k$+9) scene matrix & SUNCG & \makecell{Perceptual studies, \\Object co-occurrence map} \\
%
\midrule
\cite{wang2020sceneformer}  & Self-supervised & \makecell{2D floor image,\\ \emph{ordered} object set} & \makecell{2D CNN, \\Transformer} &  \makecell{1-channel image \\+ O$^\textbf{i}(c_i, s_i, r_i, t_i)$} & O$^\textbf{j}(c_i, s_i, r_i, t_i)$ & SUNCG & \makecell{Perceptual studies,\\ Next-object prediction accuracy} \\
%
\midrule
\cite{yang2021indoor}  & Supervised & RGB-D images & CNN-GAN & \makecell{Collection of segmented\\ depth images} & Volumetric scene (voxels) & \makecell{Structure3D, \\Matterport3D,\\ ShapeNet} & \makecell{Perceptual studies, \\Object co-occurrence map} \\
%
\midrule
\cite{paschalidou2021atiss}  & Self-supervised & \makecell{2D floor image,\\ \emph{unordered} object set} & \makecell{2D CNN,\\ Transformer} & \makecell{1-channel image \\+ O$^\textbf{i}(c_i, s_i, r_i, t_i)$} & O$^\textbf{j}(c_i, s_i, r_i, t_i)$ & 3D-FRONT & \makecell{FID score, \\Category KL divergence,\\ Real/synthetic scene classi\\-fication accuracy} \\

\bottomrule
\end{tabular*}%
}
\caption{A summary of \textbf{neural scene synthesis works} informing about the kind of learning framework employed (supervised \emph{vs.} unsupervised), the representation of indoor scenes, the kind of machinery employed to process the incorporated scene representation, the input to and output of the neural network, the dataset used and metrics employed to evaluate the generative models.}
\label{tab:doc_analysis_works}
\end{table*}

More recently, Ma et al. \cite{ma2018language} use language commands to drive scene synthesis. 
Language commands are parsed into scene graphs, which are used to retrieve subscenes from a scene database. The key idea leveraged by this method is that semantic scene graphs act as a bridge between language commands and scene arrangements, and as such, aligning the scene graph from one domain will allow retrieving corresponding scenes. To account for the lack of an exact match, the system also allows augmenting retrieved subscenes with additional objects based on the language context.  At each step, a 3D scene is synthesized by coalescing the retrieved sub-scene, with augmented objects, into the current scene.

\textbf{Human-centric scene synthesis}
Fisher et al. \cite{fisher2015activity} present a method that can produce multiple plausible 3D scenes from an input RGBD scan. here, plausibility means that the synthesized 3D scenes allow for the same functional activities as the captured environment. A scene template is estimated based on the input scan that captures likely human activities (as a probabilistic map) over the scene space. The core model, called the activity model, encodes object distribution with respect to human activities, and would guide the synthesis process based on predicted activities in the scene template.
\begin{figure*}[!t]
    \centering
    \includegraphics[width = \linewidth]{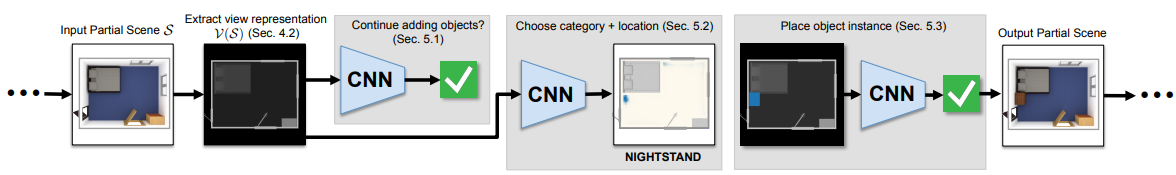}
    \caption{Scene synthesis pipeline in deep convolutional priors \cite{wang2018deep}: given an input scene, its top-down view image is obtained that contains multiple features per pixel. Such feature-rich image is analyzed using a 2D CNN to determine if an object should be added to the current scene, and if so, what type of object category and at which location. Once the category and location are determined, an instance of that category is retrieved from a model database and added to the scene at an appropriate orientation.}
    \label{fig:deep conv priors}
    \vspace{-15pt}
\end{figure*}
In a slightly different setting, Savva et al. \cite{savva2016pigraphs} capture human poses with object arrangements in the scene based on human activity. The underlying modeling is a probabilistic model. The functional relationships between humans and objects in the form of physical contacts and visual-attention linkages are represented using what they call Prototypical Interaction Graphs (in short, piGraphs). Joint probability distributions over human pose and object geometries are encoded in the PiGraphs and learned from data and in turn, the learned PiGraphs serve to guide the generation of interaction snapshots. 

A concurrent work from Ma et al. \cite{ma2016action} guides the scene generation process from human activity. In contrast to the above two methods, observations of human-object interactions in this work come from 2D images. That is, the action models are learned from annotated photographs in the Microsoft COCO dataset, which makes the problem challenging since such 2D  images do not contain object/human-pose designation. The key idea of the work is to formulate transition probabilities to account for a transition in human activity. An action graph is constructed whose nodes correspond to actions and edges encode transition probabilities. Synthesizing a new scene would correspond to sampling from the model capturing action graph priors.\\

\textbf{Deep Generative Models}
Wang et al. \cite{wang2018deep} present a deep convolutional, autoregressive approach for 3D scene synthesis. A scene is represented as a multichannel top-view image where each channel encodes the mask of an object in the scene, in addition to depth information.  An autoregressive neural network
is then trained with such images (corresponding to 3D scenes) to output object placement priors as a 2D distribution map, see Figure \ref{fig:deep conv priors}. To synthesize a new scene, objects are sequentially placed based on the learned placement priors.

\begin{figure*}[!t]
    \centering
    \includegraphics[width = \linewidth]{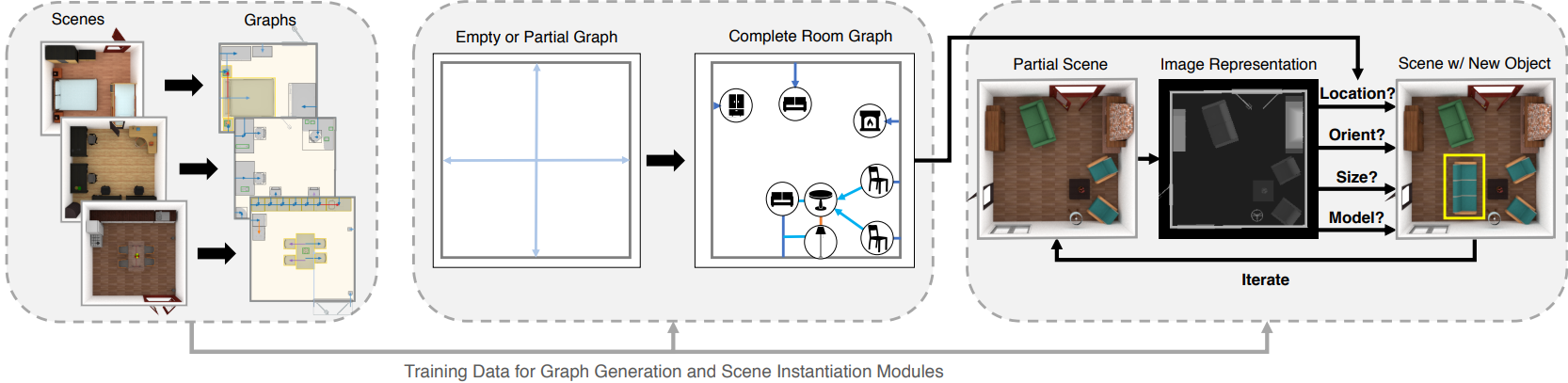}
    \caption{Scene synthesis pipeline in \emph{PlanIT} \cite{wang2019planit}: Given a database of 3D scenes, relational graphs are automatically extracted from them (left), which are fed to a graph neural network (middle). Eventually, to generate a 3D scene, a graph is instantiated using image-based reasoning (right) and 3D models corresponding to each node are inserted.}
    \label{fig:PlanIT}
\end{figure*}

Moving towards a stronger structural representation, Wang et al. \cite{wang2019planit} present an autoregressive graph generative model called \emph{PlanIT}, for 3D scenes based on Graph Neural networks employing message passing convolutions. They represent a 3D scene as a graph with scene objects as nodes and their spatial or semantic relationships by edges of a graph. During training, the network learns relationship priors between different kinds of objects in a scene type (ex: bedrooms). To generate a new scene from an empty or a partially complete scene (or scene graph), the learned autoregressive model is used to obtain a scene graph, which is instantiated via an image-based reasoning module to generate a 3D scene corresponding to that scene graph; see Figure \ref{fig:PlanIT} for an overview of this method.

\begin{figure*}[!t]
    \centering
    \includegraphics[width = \linewidth]{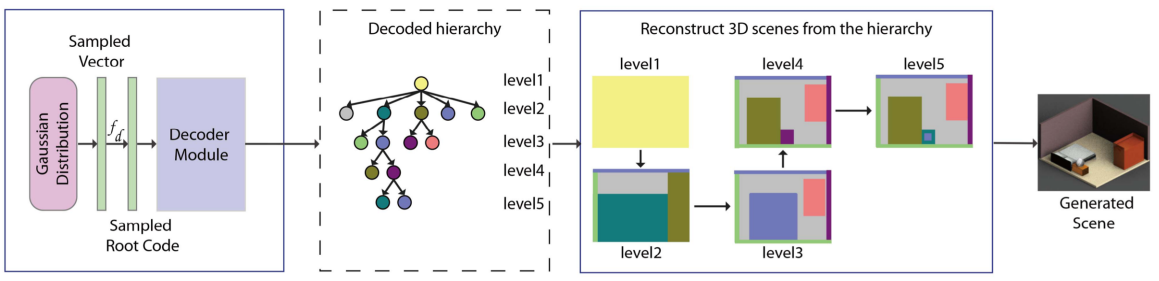}
    \caption{\emph{GRAINS} \cite{li2019grains} represents a scene as a hierarchy based on commonly occurring object relations. The learning pipeline is based on recursive neural networks (RvNN) coupled with a variational autoencoder. At inference time, a random vector is sampled from the learned latent space (which is approximated to a Gaussian distribution), and passed through the trained RvNN decoder to obtain a scene hierarchy. A 3D scene is recovered from the decoded hierarchy in a top-down fashion until all the leaf nodes of the hierarchy are traced. 3D objects are placed by retrieving from a collection of CAD models present in the scene database based on their attributes generated.}
    \label{fig:GRAINS}
    \vspace{-15pt}
\end{figure*}

Different from the above two methods, Li et al. \cite{li2019grains} present \emph{GRAINS}, a generative neural network for 3D scenes that can efficiently generate a large quantity and variety of indoor scenes. Their key observation is that indoor scenes are inherently hierarchical (so they represent 3D scenes as hierarchies), and use a recursive neural network (RvNN) architecture coupled with a VAE to model the space of scenes following the pipeline shown in Figure \ref{fig:GRAINS}. Using a dataset of annotated scene hierarchies, they train an RvNN-VAE, which performs scene
object grouping during its encoding phase and scene generation during decoding. \newline 
Specifically, a set of encoders is recursively applied to group 3D objects (represented as semantically oriented bounding boxes) in a scene, bottom up, and encodes information about the objects and their relations, where the resulting fixed-length codes roughly follow a Gaussian distribution. To generate a new scene, a random vector is sampled from the learned Gaussian and branched down through the RvNN decoder to obtain the scene hierarchy. Shape models are retrieved from a shape database based on the semantics and dimensions of leaf nodes in the generated hierarchy.\\

\begin{figure}[!t]
    \centering
    \includegraphics[width = \linewidth]{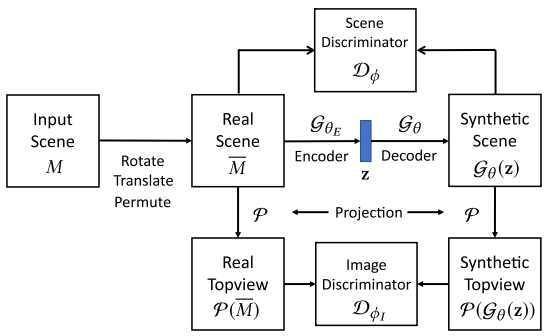}
    \caption{Schematic pipeline for scene synthesis using hybrid scene representations \cite{zhang2020deep}, where a 3D scene is represented by its top-view rendering as well as using a matrix of object properties based on their occurrences in the scene. Above, an encoder G$_{\theta_E}$ encodes a scene into a latent space, which the decoder G$_\theta$ uses to produce a scene data matrix. A scene discriminator D$_\phi$ then determines if the generated scenes are real. A projection layer $P$ projects 3D scenes to top-view images and an image discriminator D$_{\phi_I}$ classifies if its top-view images are real are not.}
    \label{fig:hybrid scene syn}
\end{figure}

\begin{figure}[!t]
    \centering
    \includegraphics[width = \linewidth]{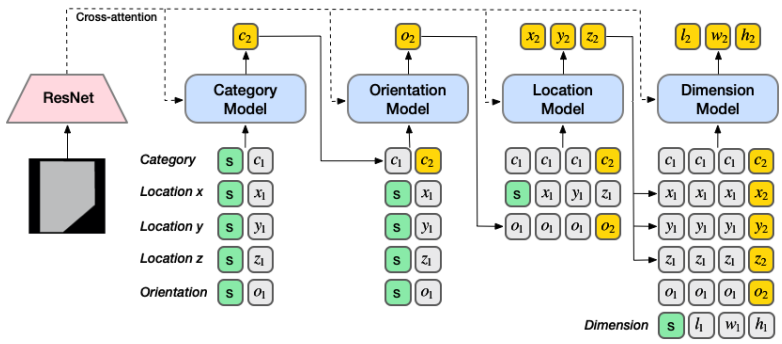}
    \caption{\emph{SceneFormer} \cite{wang2020sceneformer} uses a transformer to generate new scenes in an autoregressive manner, where a scene is represented by an ordered set of objects. In addition, it consumes a floor image, both during train and test time.}
    \label{fig:scene former}
    \vspace{-15pt}
\end{figure}

Zhang et al. \cite{zhang2020deep} present a generative model for indoor scenes based on a GAN, which learns to map a normal distribution to the distribution of primary objects in indoor scenes. In this work, a 3D scene is represented as a matrix that encodes all the information about every object in a scene. A scene is encoded into a latent vector by a set of interleaved sparse and fully connected layers. The decoder, which mirrors the encoder, generates scene matrices. A discriminator is trained to classify whether the input to it is a real scene or not. In addition, an image-based discriminator is also used to differentiate between the top-view renderings of 3D scenes; see Figure \ref{fig:hybrid scene syn} for this method's overview.\\
\newline
Very recently, \cite{wang2020sceneformer, paschalidou2021atiss} developed \emph{conditional} generative models for 3D scenes by making use of attention-based Transformer models \cite{vaswani2017attention}. The advantage of using Transformer models is that they alleviate the need for hand-crafting spatial relationships between objects, and instead, implicitly learn object relations through attention mechanism. Specifically, Wang et al. \cite{wang2020sceneformer} condition the generation process on two kinds of inputs -- room layout (including the position of doors and windows), and text descriptions. They represent indoor scenes as a sequence of object properties, converting the scene generation task to a sequence generation one. During training, an empty room (represented by the floor dimensions) or a text description (encoded using one of GloVe \cite{pennington2014glove}, ELMo \cite{peters2018deep} or BERT \cite{devlin2018bert} techniques) is input to their model along with a sequential ordering of object categories; see Figure \ref{fig:scene former} for an overall pipeline of their approach. The transformer model learns to sequentially generate the properties of the next object in the predefined ordered set. During inference, given the type of user input (empty room or text description), the trained model sequentially outputs an ordered set of objects and inserts them into the existing scene.

On the other hand, Paschalidou et al. \cite{paschalidou2021atiss} reduce the problem of scene generation to that of generating an unordered set of objects, where meaningful object arrangements are obtained by sequentially placing objects in a permutation-invariant fashion. They represent a scene as an unordered set of objects where each object is encoded using its cateory, size, orientation (relative to the floor normal), and location. During training, given a training scene with M objects, they randomly permute them and keep the first T objects (here T=3). The network is tasked to predict the next object to be added in the scene given the subset of kept objects, and the floor layout feature. During inference, they start with an empty context embedding $\textbf{C}$  and the floor representation of the room to be populated. From here, they autoregressively sample attribute values from the predicted distributions; see Figure \ref{fig:ATISS} for their method overview. Once a new object is generated, it is appended to the context $\textbf{C}$ to be used in the next step of the generation process until the end symbol is generated. To transform the predicted labeled bounding boxes to 3D models, object retrieval from the dataset based on Euclidean distance of the bounding box dimensions is performed.\\
\newline
Another recent work from Yang et al. \cite{yang2021indoor} developed a conditional volumetric generative model of indoor scenes using a GAN framework. They represent scenes as voxels, and take the room size as a conditional input to a GAN that is trained to map the distribution of an indoor scene to a normal distribution. The discriminator is trained on depth and semantic images of the volumetric scenes. To this end, they employ a differentiable renderer to render depth and semantic maps of generated volumetric scenes, which are used with the depth, and semantic maps of scenes from the training database for learning the GAN discriminator. 
At generation time, given a room size $\phi$ and a latent vector \textbf{zs} randomly
sampled from the latent space, the trained volumetric GAN can generate a semantic scene volume that stores both layout and rough shapes of the objects instances in the room. To obtain the final 3D scene, they extract object instances from the semantic scene volume and replace them with the CAD models retrieved (based on Chamfer Distance) from a 3D object database.
%

\textbf{Discussion}
With the availability of large synthetic 3D scene datasets such as 3D-FRONT \cite{fu20213d}, and the impressive advancements made in developing generative neural networks for 3D scenes, a basic question naturally arises -- do we need more such generative models, and what purpose would more of such scenes serve anyway?

While it is surely worth having access to large quantities of synthetic and generated 3D scenes, they are not of practical use unless they are functional. That is, human activity should be adequately supported by these scenes. For example, if the area for in-and-out movement for a family of four in a generated living room is insufficient, then such a generated scene, although appearing plausible, does not find real-world applicability. This issue is systemic, in the sense that the typical way to scene synthesis has been to treat humans and scenes separately. We need to model them together, allowing us to generate functionally plausible environments, and in turn, using such scenes to improve human pose within that space and optimize activity. Scenes and humans complement each other.


\revision{
In recent years, neural scene rendering \cite{mildenhall2021nerf}, diffusion models \cite{sohl2015deep, ho2020denoising}  and CLIP-based models \cite{radford2021learning} have gained a lot of traction for generating novel data samples. They have been predominantly employed in the 2D domain, with only recent exploration for 2D-to-3D generation \cite{gao2022get3d, poole2022dreamfusion} on single-object images. Leveraging these models for 3D indoor scenes is an under-explored direction mainly due to the complexity of the scene structures. Below, we present recent works that use these newer techniques for synthesizing indoor scenes. 

\begin{figure}[!t]
    \centering
    \includegraphics[width = \linewidth]{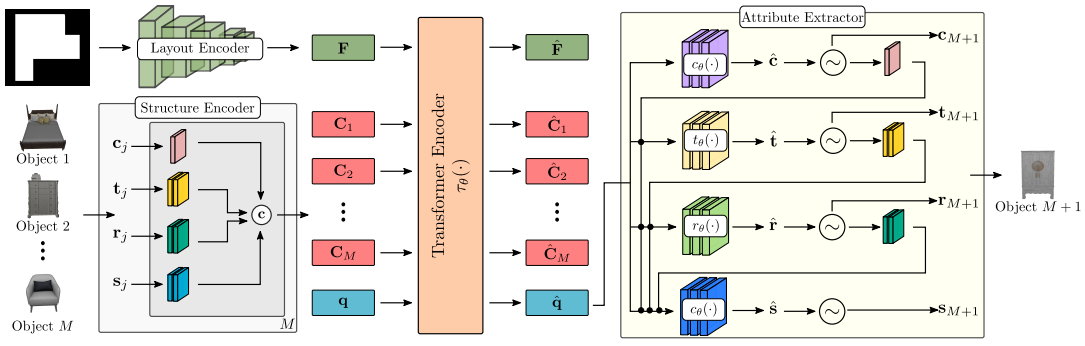}
    \caption{\emph{ATISS} \cite{paschalidou2021atiss} poses the scene synthesis task as one of unordered set generation. Given a room type and its shape (in the form of a top-down floor image), it generates plausible furniture arrangements in an autoregressive, permutation-invariant fashion using a transformer.}
    \label{fig:ATISS}
    \vspace{-15pt}
\end{figure}



\textbf{Using Radiance Fields}
The basic idea in neural rendering is to first sample spatio-temporal coordinates with respect to a given 3D scene, feed it through a neural network to recover radiance/signed-distance fields and employ a differentiable forward map (such as sphere tracing or volume rendering) that outputs a novel view RGB image (see \cite{xie2022neural} for a detailed report).
Using these radiance fields to hallucinate scene geometry or to manipulate object arrangements allows their employability for indoor scene modeling applications. 


To this end, \cite{devries2021unconstrained, yang2021learning} present some of the early works on using radiance fields for scene hallucination/completion and scene editing, respectively. Yang et al. \cite{yang2021learning} develop a neural rendering system that enables editing on real-world scenes by learning an object-compositional neural radiance field. The main idea is to use two separate branches to encode the scene, one to process the scene background and the other to process the constituent objects, both undergoing a neural rendering pipeline of their own. In doing so, the system runs a neural radiance field pipeline on the object branch that takes as inputs object voxel features and an object activation code, allowing scene edits at the object level based on the object activation code.

DeVries et al. \cite{devries2021unconstrained} make use of the radiance fields to learn scene priors from which novel scenes can be sampled. Specifically, they train a GAN in which the generator learns to decompose scenes into a collection of many \emph{local} radiance fields that can be rendered from a freely moving camera. The generator of this proposed GAN tries to learn a distribution of novel view images (using a NeRF model) that is similar to the prior distribution. The discriminator is tasked to classify the images at the output of generator to be fake. Once trained on many diverse scenes and viewpoints, a novel scene sample can be generated using a random vector that can hallucinate parts of a scene captured in the training images.

\cite{wang2022neuralroom} propose NeuralRooms to reconstruct indoor scenes (represented by meshes) from unposed multi-view 2D images based on neural radiance fields. The main motivation of this work is that shape-radiance ambiguity and the presence of texture-less regions in indoor scenes make it difficult to faithfully reconstruct them using multi-view stereo (MVS) algorithms or using NeRF models. To address this issue, they propose a two-part learning framework wherein the first part makes use of a MVS algorithm \cite{schonberger2016structure} (ensuring the accuracy of texture-rich and edge areas) and a normal estimation network \cite{bae2021estimating} (ensuring completeness of texture-less regions) to acquire geometry prior. The input RGB images and the geometry prior are used in a neural rendering-based surface reconstruction pipeline. Finally, ray-tracing on the reconstructed scene is done with TSDF fusion to obtain a 3D mesh for the reconstructed indoor scene.

\begin{figure*}[!t]
    \centering
    \includegraphics[width = \linewidth]{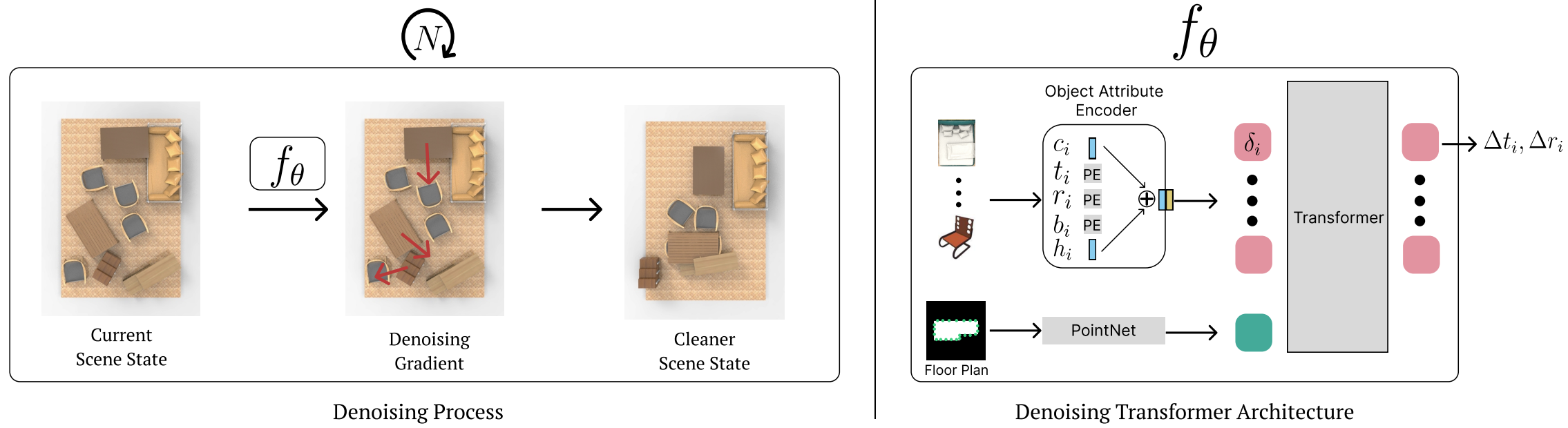}
    \caption{Denoising process of LEGO-Net \cite{wei2023lego}(left) and the backbone transformer (right) performing the denoising process. LEGO-Net takes the current scene state and iteratively modifies the scene state by computing denoising gradient towards the clean manifold. The transformer on the right transformer takes the scene attributes of the current state and outputs 2D transformations of each object.}
    \label{fig:lego-net}
    \vspace{-15pt}
\end{figure*}

\textbf{Using Diffusion and CLIP-based models}
Unlike in the 2D domain, the deployment of diffusion models for modeling 3D indoor scenes has not seen much activity. To the best of our knowledge, the recent works of Lego-Net \cite{wei2023lego} and \cite{lei2022generative} are the only work devoted to this task. 

Lego-Net \cite{wei2023lego} develops a denoising diffusion model to learn the rearrangement of objects in a messy indoor scene, where, similar to ATISS, a scene is represented by an unordered set of objects and their transformations. The underlying model that is used is a transformer. Given an input messy scene, the transformer iteratively computes the denoising gradient towards the clean manifold, until a so-called regular state is reached; see Figure \ref{fig:lego-net}. Note that the denoising process is not driven by any end goal state, and the concept of an end goal state is not made use of in this work. At each step of the denoising process, the transformer takes scene attributes of the current state and outputs 2D transformations of each object that would make the scene ``cleaner". Training this model takes place in a reverse fashion where clean scenes are perturbed to make them messy. 

Lei et al. \cite{lei2022generative}, on the other hand, propose to synthesize an indoor scene via incremental inpainting. They represent indoor scenes as RGBD images. Given a sparse set of multi-view RGBD images, the goal is to generate a coherent 3D scene mesh by predicting RGBD frames along a \emph{novel} camera trajectory. To this end, they first fuse the input view images into an initial mesh, which is then rendered to get an incomplete, hole-present image of the scene. This incomplete scene is then inpainted using a RGBD diffusion model which is back-projected to the 3D domain to get a 3D mesh. This mesh is integrated with the very initial mesh to get a new mesh. This process takes place in an iterative fashion, eventually generating a novel indoor scene.

CLIP \cite{radford2021learning} is a popular text-to-image generative model capable of producing novel 2D images from text prompts. In the 3D domain, CLIP has been used to generate 3D shapes as shown in \cite{sanghi2022clip, sanghi2022textcraft}. Its extension to generating 3D indoor scenes is not straightforward owing to the presence of multiple objects. CLIP-Layout \cite{liu2023clip} presents the first work that leverages a CLIP model to synthesize 3D indoor scenes. 

Similar to ATISS \cite{paschalidou2021atiss}, CLIP-Layout \cite{liu2023clip} adopts an auto-regressive approach for indoor 3D scene synthesis, but the synthesis is now additionally constrained on a text input that acts as a style description prompt. This work also encodes a 3D scene as an unordered set of objects and their transformations, and the underlying machinery is based on a transformer model. The text prompt is encoded using a CLIP encoder \cite{radford2021learning}, which takes eight view-renderings of a 3D scene as input and outputs a 512-dimensional vector, which is then concatenated with all other object-plus-floor features extracted using the transformer as done in ATISS. Note that a pre-trained text-to-image CLIP encoder is used during the training phase. This allows the representation of fine visual details of each furniture instance while remaining agnostic to object encoding format. Sample results are shown in Figure \ref{fig:clip_layout}.

\begin{figure}[!t]
    \centering
    \includegraphics[width = \linewidth]{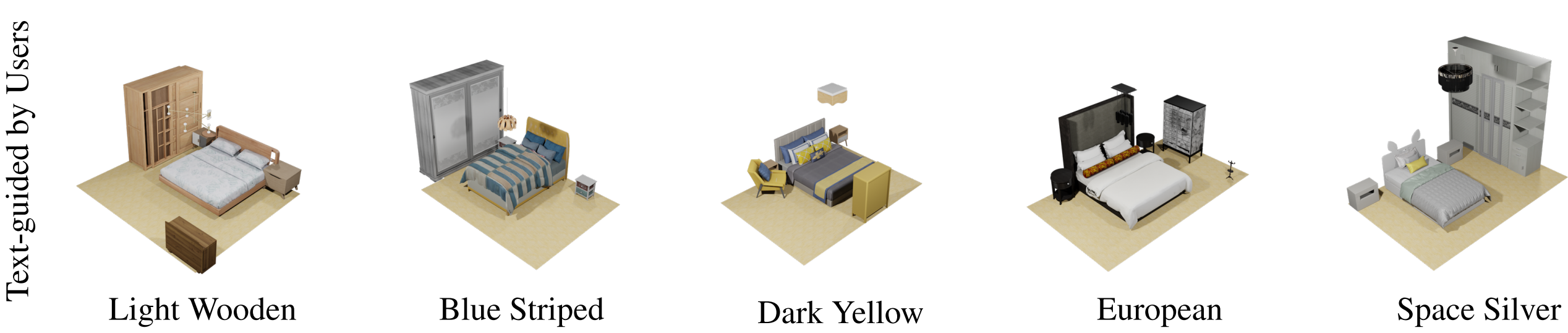}
    \caption{CLIP-Layout \cite{liu2023clip} takes a floorplan and a text prompt as conditions for style descriptions, and generates plausible, diverse 3D indoor scenes that are stylistically consistent, adhering to the input text prompt.}
    \label{fig:clip_layout}
    \vspace{-15pt}
\end{figure}
}
\section{Conclusions and open problems}
\label{sec:open_research_problems}

In this report, we have surveyed historical and state-of-the-art works in data-driven analysis as well as synthesis of 3D indoor scenes. Starting from different possible representations (both visual and structural) and the available datasets, we discussed fundamental scene analysis tasks such as 3D object detection, 3D scene segmentation, 3D scene reconstruction, and 3D scene retrieval. For synthesis techniques, we have mainly documented recent progress in that direction, which by default, has been skewed towards neural models. During the course of these discussions, we have identified the suitability of specific neural architecture for the chosen scene representations. 

Overall, indoor scene modeling has made impressive strides in recent years, pushing the boundaries of computer graphics research. Yet, there exist many interesting avenues at a higher level to be pursued. We conclude this report by offering thoughts on what we regard as some important and interesting research directions.

\textbf{Modeling rotation equivariance for 3D object detection: }
A strongly desired property in 3D deep learning is rotation equivariance. In the context of object detection in 3D indoor scenes, it is desired that the detected bounding box be \emph{equivariant} to the object pose. This means achieving equivariance not at the global input scene level, but at the object level. The first step towards this goal is to extract equivariant features at the object level, and finding a way to inject them into recent 3D object detectors. Since objects in a scene are rotated along the gravity axis, it is natural to limit the development of techniques to 1D rotations. A recent work of \cite{yu2022rotationally} has laid some groundwork in this regard. However, it is the equivariance in SO(3) space that is challenging and under-explored, finding wide industrial applicability such as in simulating flight aerodynamics, building constructions, and in assembly lines in manufacturing industries.

\textbf{Instance detection and segmentation in scenes: }
In the real world, sets of \emph{identical} objects are observed in different orientations. In synthetic scenes, such identical objects are typically represented with instances of the \emph{same} 3D model appropriately transformed into various positions. The ability to segment scenes, not simply semantically, but also at the instance level, provides a greater machine understanding of indoor environments. The solution to this problem raises a more fundamental question that is related to modeling object rotations in the SO(3) space -- can existing shape descriptors distinguish between rotated instances of the same object, and if not, how can we go about doing that? In addition, if an agent can identify instances of the same objects in a scene (through instance segmentation), motion articulations on one such model can be easily applied to all other instances of the model in the scene. These are all connected problems, but can only be attempted if the fundamental question posed above has been firmly answered. 

\textbf{Unsupervised scene reconstruction from a single image: }
Object reconstruction from single-view images is a difficult problem in itself. At the scene layout level, different challenges exist, particularly of cluttered scene recovery. The development of such models can facilitate visualizing possibilities for architects, furniture retailers, and interior design firms. Recent supervised works \cite{nie2020total3dunderstanding, zhang2021holistic} have achieved decent results on this task while also performing reconstruction at the object level. When no layout supervision is provided, the problem of recovering the scene layout from a single-view image becomes extremely challenging. A potential solution to this problem is via self-supervision, where the model needs to reason about placement priors from many homogeneous scene images. A lack of existing works in this direction indicates the degree of difficulty involved in the task, something to actively research about.

\textbf{Neural 3D scene similarity: }
Alignment of visual data provides a reference point to study data similarity. For 3D scenes, this is notoriously hard, since there is no standard reference point -- it all depends on the scene object in focus. Developing algorithms for 3D scene similarity has continued to attract interest, starting from graph kernels \cite{fisher2011characterizing} to later works on focal-centric graph kernels \cite{xu2014organizing} and IBS \cite{zhao2014indexing}. A recent work \cite{patil2021layoutgmn} uses graph neural network (GNN) to learn structural similarity for 2D layouts. To explore a deployment of GNNs with intuitive focal-centric approaches that combine not just the structural layout properties, but also object appearance traits, for 3D scene similarity/retrieval remains an interesting direction. Such a metric could also be used to qualitatively evaluate the plausibility of 3D scenes, which is a missing piece in the literature, especially in the realm of neural generative models.

\textbf{Modeling geometry and object textures for scene synthesis: }
Access to a rich 3D scene database that contains diverse 3D objects is crucial for building neural models tasked with scene generation. In all such works, objects are placed in the generated scene layout by retrieving from an object database based on generated object attributes, which typically are nothing but bounding box dimensions, category, and scene centroid. Although geometric information is generated, appearance properties are not accounted for. This needs to be addressed since existing structural and geometric attributes provide strong cues to material and texture appearance, something that has been under-explored at the object level \cite{jain2012material, lin2018learning, chen2015magic}. Learning to model a coupling of this with the scene layouts is an interesting approach to bypass the typical object retrieval step and directly generate objects with novel appearances.

\textbf{Neural text-to-3D scene: }
With DALL-E \cite{ramesh2022hierarchical} and IMAGEN \cite{saharia2022photorealistic}, great advances have been made this year for language-driven image synthesis that produce impressive, high-quality results. However, these models do not offer control over the generated results. That is, the underlying theme is of a one-shot text-conditioned generation, something that is rarely desired when dealing with 3D content. Rather, the ability to progressively generate 3D content in a controlled manner is prioritized in the graphics community. In the realm of indoor 3D scenes, a few attempts \cite{chang2014learning, chang2015shapenet, ma2018language} have been made in the past to generate scenes from text input. All these works are model-driven, and are limited by incorporated heuristics. Admitting neural networks for this task that learn directly from the data can make up for the ``lost ground", and open up further avenues to improve such scene generation systems. 

\textbf{Modeling scene style: }
Scene style is a high-level concept, generally referring to a setting where the furniture designs (ornate antique sofa vs. a flat IKEA sofa) are in agreement with, and even complement, the ambient decorations in the scene (ex; ceiling arches, wall carvings and other beautification), and themselves exhibit geometric uniformity. In addition, room color, ambiance lighting, furniture color, texture and material, all factor into what we call scene style. To be able to generate scenes of different styles implies modeling a function of object styles and room decorations. A recent work \cite{solah2022mood} gives a teaser on scene style, based on a person's mood. The question of how to define this function and how to factorize object/room style in terms of material, color and texture remains an open research problem.

In addition to such fine details, scene ``style" can also refer to its tidiness. Real world scenes are often cluttered with objects and are far from the impeccable renderings depicted in the literature. A significant energy is devoted to producing near-perfect scenes, ignoring investigations into producing messy (sub)scenes that mimic environments encountered in daily life. Learning to model such scenes, although challenging, can be beneficial in providing realism to virtual scenes and remains an open research direction.

\textbf{Modeling scene interaction: }
For an immersive metaverse experience, users should be able to interact with objects in a scene. This can occur in two ways -- (1) using objects and furniture in their intended purposes (ex: sitting on a chair, adjusting the sofa to appropriately face the tv, using the tv remote, picking up a pen), and (2) playing with furniture models (ex: facilitating articulations on the drawers of a file cabinet, where the user can touch a drawer and it pops open; allowing part mobility such as adjusting the height of a chair seat; manipulating object functionalities in novel ways, perhaps making them non-functional). There have been some attempts in modeling object functionalities and part mobility \cite{sharf2014mobility, li2015foldabilizing, hu2015interaction, hu2016learning, hu2017learning, hu2020predictive}. However, modeling interactions at the scene level poses novel challenges and is an interesting research direction that has been under-explored.

\textbf{Move planning, scene rearrangement, and teleportation: }
The ability to move freely and efficiently in indoor environments influences productivity and work culture, either for shared workplaces such as offices, restaurant kitchens, warehouses, and airports or for personal spaces such as living rooms and open kitchens. More often than not, such planning takes ample time and layout considerations, especially with a large object inventory. Suggesting a plausible arrangement of objects leading to an optimized workplan, as well as space design is extremely useful in industrial applications. In a recent work \cite{zhang2021joint}, such workspaces and workplans are automatically designed given the input space and workspace equipment, in addition to staff properties as inputs. Such designs may benefit from data-driven modeling, for which, a diverse, rich, large-scale database of indoor environments spanning different industries is the first necessary step. 

This move planning concept can be used to inject teleportation feature in the metaverse, where a system can automatically suggest possible teleportation locations within an already-seen environment based on navigability, and the user can hop between such spots virtually. One way to do this is to sample a set of desirable teleport positions, assessing ease of navigation (and properties such as coverage and connectivity from a subscene or a focal point). Such an application has recently been explored in \cite{li2021synthesizing} which synthesizes scene-aware teleportation graphs.

A slight deviation from the above, but with the potential to serve teleportation application, albeit slower, would be to allow the user to select teleportation spots apriori in an already configured environment, and developing a system that can re-arrange objects in the current physical state of the environment to a new state so as to optimize for the desired tele-movement. Different versions of this task exist have been described in \cite{batra2020rearrangement} where the target environment state can be described by object poses, images, language description, or by letting an agent experience the target state environment, if possible. A recent work \cite{wang2020scene} makes use of reinforcement learning for automatic move planning of 3D objects from an initial 3D layout to a target layout. This application serves well in practice, and requires further exploration.

\section{Author Bios}

\textbf{\href{http://www.sfu.ca/~agadipat/}{Akshay Gadi Patil}}
obtained his Ph.D. from Simon Fraser University, Canada, working in the GrUVi lab. His research focuses on understanding and modeling 3D shapes and scenes using machine learning techniques, which is strongly related to the theme of this report. With his collaborators, he developed the first hierarchical deep learning framework for scene synthesis, named \href{https://dl.acm.org/doi/10.1145/3303766}{GRAINS}, and also authored a SIGGRAPH Asia paper on synthesizing 3D scenes using compact natural language. His work on 2D layouts, named \href{https://openaccess.thecvf.com/content_CVPRW_2020/papers/w34/Patil_READ_Recursive_Autoencoders_for_Document_Layout_Generation_CVPRW_2020_paper.pdf}{READ}, won the best paper award at \href{https://cvpr2020text.wordpress.com/}{this} CVPR 2020 workshop. Earlier, he earned a M.Tech degree in Electrical Engineering from the Indian Institute of Technology Gandhinagar. He co-organized the \href{https://learn3dg.github.io/}{Learning to Generate 3D Shapes and Scenes} workshop at ECCV 2022.

\noindent \textbf{\href{https://supriya-gdptl.github.io/}{Supriya Gadi Patil}}
earned her M.Sc. degree from Simon Fraser University, Canada. She was a part of the GrUVi lab, where she worked on incorporating deep learning techniques for geometric modeling of 3D shapes and scenes, mainly in the realm of reconstruction. Her expertise includes domain knowledge of various 3D indoor scene datasets and scene reconstruction methods, which nicely complements the content delivery of this survey report. In the past, she earned a M.Tech degree in Computer Science from the Indian Institute of Technology Hyderabad working in the area of Graph Neural Networks, and was a student researcher at the Max Plank Institute for Software Systems (MPI-SWS), Germany. In the industry, she worked on integrating machine learning models into commercial products as a full-time employee at Adobe India.

\noindent \textbf{\href{https://manyili12345.github.io/}{Manyi Li}} 
is an Associate Researcher in the School of Software at Shandong University. She received B.Sc. and Ph.D. degrees from Shandong University in 2013 and 2018 respectively and was a Post-Doctoral research scholar in the GrUVi Lab at Simon Fraser University, from 2019–2021. Her main interests are in 3D content creation and understanding, with a special focus on 3D man-made objects and indoor scenes, which is highly correlated with this report. With her collaborators, she authored the first hierarchical deep generative model, named \href{https://dl.acm.org/doi/10.1145/3303766}{GRAINS}, catered towards fast and diverse synthesis of 3D indoor scenes. She has also worked on developing unsupervised learning techniques for object understanding and reconstruction, 
which are relevant to object insertion in 3D indoor scenes. She was a co-organizer of the \href{https://learn3dg.github.io//static/2021.html}{Learning to Generate 3D Shapes and Scenes} workshop at CVPR 2021.

\noindent \textbf{\href{https://techmatt.github.io/}{Matthew Fisher}}
Matthew Fisher is a Principal Scientist at Adobe Research. He obtained his Ph.D. in Computer Science from Stanford University and a B.S. in CS from the California Institute of Technology. Matt's research focuses on combining computer graphics, vision, and machine learning to make it faster and more fun to complete creative tasks. He has published over 50 papers in graphics and vision, including many foundational papers in scene understanding and scene synthesis such as \href{https://techmatt.github.io/pdfs/graphKernel.pdf}{3D Graph Kernel}, \href{https://graphics.stanford.edu/projects/scenesynth/}{SceneSynth}, \href{https://techmatt.github.io/pdfs/actSynth.pdf}{FunctionalSceneSynth} and his Ph.D. thesis was on data-driven tools for scene modeling. His recent work looks into applying deep generative models to help artists more quickly design and model 3D objects and scenes. At Adobe Research, he works extensively with creative professionals to build and deliver new tools that accelerate the artistic process across painting, animation, and video editing applications.

\noindent \textbf{\href{https://msavva.github.io/}{Manolis Savva}}
is an Assistant Professor at Simon Fraser University, and a Canada Research Chair in Computer Graphics. He completed his PhD at the Stanford graphics lab, advised by Pat Hanrahan. His research focuses on human-centric 3D scene analysis, 3D scene generation, and simulation for scene understanding. He has also worked in data visualization, grounding of natural language to 3D content, and in creating large-scale datasets for 3D deep learning. He has published a number of papers in related areas including \href{https://3dlg-hcvc.github.io/plan2scene/}{Plan2Scene}, \href{https://aihabitat.org/}{Habitat}, and \href{https://graphics.stanford.edu/projects/pigraphs/}{PiGraphs}. He organized workshops on relevant topics such as the \href{https://embodied-ai.org/}{Embodied AI} workshop (CVPR 2020, CVPR 2021) and \href{https://shapenet.cs.stanford.edu/iccv17workshop/}{Learning to See from 3D Data} (ICCV 17). He has also taught a related course \href{https://dl.acm.org/doi/10.1145/3415263.3419152}{Learning 3D Functionality Representations} at SIGGRAPH Asia 2020, and co-authored and presented a STAR on \href{https://onlinelibrary.wiley.com/doi/abs/10.1111/cgf.13385}{Functionality Representations and Applications for Shape Analysis} at EuroGraphics 2018.

\noindent \textbf{\href{https://www2.cs.sfu.ca/~haoz/}{Hao (Richard) Zhang}}
is a professor in the School of Computing Science at Simon Fraser University, Canada, and an Amazon Scholar. He obtained his Ph.D. from the Dynamic Graphics Project (DGP), University of Toronto, and M.Math. and B.Math degrees from the University of Waterloo, all in computer science. Richard's research is in computer graphics with special interests in geometric modeling, analysis and synthesis of 3D contents (e.g., shapes and indoor scenes), geometric deep learning, as well as computational design and fabrication. He has published more than 170 papers on these topics. Most relevant to this survey, Richard was one of the co-authors of the Eurographics STARs on \href{https://www.cs.sfu.ca/~haoz/pubs/mitra_star13.pdf}{Structure-Aware Shape Processing}, \href{https://www.cs.sfu.ca/~haoz/pubs/egstar2020.pdf}{Learning Generative Models of 3D Structures}, and taught SIGGRAPH courses on closely related topics such as "modeling and remodeling 3D worlds." With his collaborators, he has made original and impactful contributions to structural analysis and synthesis of 3D shapes and environments including co-analysis, hierarchical modeling, semi-supervised learning, topology-varying shape correspondence and modeling, and deep generative models.

\bibliography{9-references}
\bibliographystyle{eg-alpha.bst}

\end{document}